\documentclass{statsoc}
\usepackage{natbib}
\usepackage{amsmath,amssymb}
\usepackage{graphicx}
\usepackage{ulem}
\usepackage{times}
\usepackage{color}
\usepackage{anysize}

\marginsize{1in}{1in}{0.8in}{0.65in}

\newtheorem{condition}{Condition}
\newtheorem{theorem}{Theorem}
\newtheorem{lemma}{Lemma}

\newtheorem{definition}{Definition}

\newcommand\relph[1]{\mathrel{\phantom{#1}}}
\DeclareMathOperator*\argmin{\arg\min}
\DeclareMathOperator\diag{diag}
\DeclareMathOperator\tr{tr}
\DeclareMathOperator\unif{unif}
\DeclareMathOperator\rep{rep}

\def\bA{\mathbf{A}}
\def\bB{\mathbf{B}}
\def\bC{\mathbf{C}}
\def\bD{\mathbf{D}}
\def\bE{\mathbf{E}}
\def\bF{\mathbf{F}}
\def\bG{\mathbf{G}}
\def\bH{\mathbf{H}}
\def\bI{\mathbf{I}}
\def\bK{\mathbf{K}}
\def\bL{\mathbf{L}}
\def\bM{\mathbf{M}}
\def\bU{\mathbf{U}}
\def\bV{\mathbf{V}}
\def\bW{\mathbf{W}}
\def\bX{\mathbf{X}}
\def\bY{\mathbf{Y}}
\def\bZ{\mathbf{Z}}
\def\ba{\mathbf{a}}
\def\bb{\mathbf{b}}
\def\be{\mathbf{e}}
\def\bff{\mathbf{f}}
\def\bg{\mathbf{g}}
\def\bm{\mathbf{m}}
\def\bu{\mathbf{u}}
\def\bv{\mathbf{v}}
\def\bw{\mathbf{w}}
\def\bx{\mathbf{x}}
\def\tbx{\widetilde{\bx}}
\def\by{\mathbf{y}}
\def\cA{\mathcal{A}}
\def\cB{\mathcal{B}}
\def\cC{\mathcal{C}}
\def\cK{\mathcal{K}}
\def\cS{\mathcal{S}}
\def\bGamma{\boldsymbol{\Gamma}}
\def\bbeta{\boldsymbol{\beta}}
\def\bDelta{\boldsymbol{\Delta}}
\def\bLambda{\boldsymbol{\Lambda}}
\def\bTheta{\boldsymbol{\Theta}}
\def\bOmega{\boldsymbol{\Omega}}
\def\bSigma{\boldsymbol{\Sigma}}
\def\blambda{\boldsymbol{\lambda}}
\def\ve{\varepsilon}
\def\bve{\boldsymbol{\varepsilon}}
\def\bzero{\mathbf{0}}
\def\bone{\mathbf{1}}
\def\what{\widehat}
\def\wtilde{\widetilde}
\newcommand{\rank}{\mathop{\rm rank}\nolimits}
\DeclareMathOperator\sgn{sgn}
\DeclareMathOperator\vect{vec}

\title[SOFAR
]{SOFAR: large-scale association network learning
\thanks{This work was supported by a Grant-in-Aid for JSPS Fellows 26-1905, NSF CAREER Awards DMS-0955316 and DMS-1150318, 
NIH 
grant U01 HL114494, NSF Grant DMS-1613295, 
and a grant from the Simons Foundation. The authors would like to thank the Joint Editor, Associate Editor, and referees for their valuable comments that have helped improve the paper significantly. Part of this work was completed while Fan and Lv visited the Departments of Statistics at University of California, Berkeley and Stanford University. These authors sincerely thank both departments for their hospitality.}}
\author{Yoshimasa Uematsu$^1$, Yingying Fan$^1$, Kun Chen$^2$, Jinchi Lv$^1$ and Wei Lin$^3$}
\address{University of Southern California$^1$, University of Connecticut$^2$ and Peking University$^3$}

\begin{document}
\begin{abstract}
Many modern big data applications feature large scale in both numbers of responses and predictors. Better statistical efficiency and scientific insights can be enabled by understanding the large-scale response-predictor association network structures via layers of sparse latent factors ranked by importance. Yet sparsity and orthogonality have been two largely incompatible goals. To accommodate both features, in this paper we suggest the method of sparse orthogonal factor regression (SOFAR) via the sparse singular value decomposition with orthogonality constrained optimization to learn the underlying association networks, with broad applications to both unsupervised and supervised learning tasks such as biclustering with sparse singular value decomposition, sparse principal component analysis, sparse factor analysis, and spare vector autoregression analysis. Exploiting the framework of convexity-assisted nonconvex optimization, we derive nonasymptotic error bounds for the suggested procedure characterizing the theoretical advantages. The statistical guarantees are powered by an efficient SOFAR algorithm with convergence property. Both computational and theoretical advantages of our procedure are demonstrated with several simulation and real data examples.
\end{abstract}

\quad
\vspace{-0.19in}



\section{Introduction} \label{Sec1}


The genetics of gene expression variation may be complex due to the presence of both local and distant genetic effects and shared genetic components across multiple genes \citep{Brem:Krug:land:2005,Cai:Li:Liu:Xie:cova:2013}. A useful statistical analysis in such studies is to simultaneously classify the genetic variants and gene expressions into groups that are associated. For example, in a yeast expression quantitative trait loci (eQTLs) mapping analysis, the goal is to understand how the eQTLs, which are regions of the genome containing DNA sequence variants, influence the expression level of genes in the yeast MAPK signaling pathways. Extensive genetic and biochemical analysis has revealed that there are a few functionally distinct signaling pathways of genes \citep{Gustin1998, Brem:Krug:land:2005}, suggesting that the association structure between the eQTLs and the genes is of low rank. Each signaling pathway involves only a subset of genes, which are regulated by only a few genetic variants, suggesting that each association between the eQTLs and the genes is sparse in both the input and the output (or in both the responses and the predictors), and the pattern of sparsity should be pathway specific. Moreover, it is known that the yeast MAPK pathways regulate and interact with each other \citep{Gustin1998}. The complex genetic structures described above clearly call for a joint statistical analysis that can reveal multiple distinct associations between subsets of genes and subsets of genetic variants. 
If we treat the genetic variants and gene expressions as the predictors and responses, respectively, in a multivariate regression model, the task can then be carried out by seeking a sparse representation of the coefficient matrix and performing predictor and response selection simultaneously. The problem of large-scale response-predictor association network learning is indeed of fundamental importance in many modern big data applications featuring large scale in both numbers of responses and predictors.

Observing $n$ independent pairs $(\bx_i, \by_i)$, $i=1,\cdots, n$, with $\bx_i\in \mathbb{R}^p$ the covariate vector and $\by_i\in\mathbb{R}^q$ the response vector, motivated from the above applications we consider the following multivariate regression model
 \begin{align}
\bY=\bX \bC^*+\bE, \label{eq:model}
 \end{align}
 where $\bY=(\by_1,\dots,\by_n)^T\in\mathbb{R}^{n\times q}$ is the response matrix, $\bX=(\bx_1,\dots,\bx_n)^T\in\mathbb{R}^{n\times p}$ is the predictor matrix, $\bC^*\in\mathbb{R}^{p\times q}$ is the true regression coefficient matrix, and $\bE=(\be_1,\dots,\be_n)^T$ is the error matrix.   To model the sparse relationship between the responses and the predictors as in the yeast eQTLs mapping analysis, we exploit the following singular value decomposition (SVD) of the coefficient matrix
 \begin{align}
& \bC^*=\bU^*\bD^*\bV^{*T}=\sum_{j=1}^rd_j^*\bu_j^*\bv_j^{*T},  \label{eq:svd}
 \end{align}
 where $1\le r\le\min(p,q)$ is the rank of matrix $\bC^*$, $\bD^*=\diag(d_1^*,\dots,d_r^*)$ is a diagonal matrix of nonzero singular values, and $\bU^*=(\bu_1^*,\dots,\bu_r^*)\in\mathbb{R}^{p\times r}$ and $\bV^*=(\bv_1^*,\dots,\bv_r^*)\in\mathbb{R}^{q\times r}$ are the orthonormal matrices of left and right singular vectors, respectively. Here, we assume that $\bC^*$ is low-rank with only $r$ nonzero singular values, and the matrices $\bU^*$ and $\bV^*$ are sparse. 

Under the sparse SVD structure \eqref{eq:svd}, model \eqref{eq:model} can be rewritten as
\[
\wtilde\bY=\wtilde\bX\bD^*+\wtilde\bE,
\]
where $\wtilde\bY=\bY\bV^*$, $\wtilde\bX=\bX\bU^*$, and $\wtilde\bE=\bE\bV^*\in\mathbb{R}^{n\times r}$ are the matrices of latent responses, predictors, and random errors, respectively. The associations between the predictors and responses are thus diagonalized under the pairs of transformations specified by $\bU^*$ and $\bV^*$. When $\bC^*$ is of low rank, this provides an appealing \textit{low-dimensional latent model} interpretation for model \eqref{eq:model}. Further, note that the latent responses and predictors are linear combinations of the original responses and predictors, respectively. Thus, the interpretability of the SVD can be enhanced if we require that the left and right singular vectors be  sparse so that each latent predictor/response involves only a small number of the original predictors/responses, thereby performing the task of variable selection among the predictors/responses, as needed in the yeast eQTLs analysis.

The above model \eqref{eq:model} with low-rank coefficient matrix has been commonly adopted in the literature. In particular, the reduced rank regression \citep{Ande:esti:1951,Izen:redu:1975,Rein:Velu:mult:1998} is an effective approach to dimension reduction by constraining the coefficient matrix $\bC^*$ to be of low rank. \citet{Bune:She:Wegk:opti:2011} proposed a rank selection criterion that can be viewed as an $L_0$ regularization on the singular values of $\bC^*$. The popularity of $L_1$ regularization methods such as the Lasso \citep{Tibs:regr:1996} led to the development of nuclear norm regularization in multivariate regression \citep{Yuan:Ekic:Lu:Mont:dime:2007}. 
\citet{Chen:Dong:Chan:redu:2013} proposed an adaptive nuclear norm penalization approach to bridge the gap between $L_0$ and $L_1$ regularization methods and combine some of their advantages. With the additional SVD structure \eqref{eq:svd},  \citet{Chen:Chan:Sten:redu:2012} proposed a new estimation method with a correctly specified rank by imposing a weighted $L_1$ penalty on each rank-1 SVD layer for the classical setting of fixed dimensionality. \citet{Chen:Huan:spar:2012} and \citet{Bune:She:Wegk:join:2012} explored a low-rank representation of $\bC^*$  in which the rows of $\bC^*$ are sparse; however, their approaches do not impose sparsity on the right singular vectors and, hence, are inapplicable to settings with high-dimensional responses where response selection is highly desirable.

Recently, there have been some new developments in sparse and low-rank regression problems. \citet{MaSun2014} studied the properties of row-sparse reduced-rank regression model with nonconvex sparsity-inducing penalties, and later \citet{MaSun2016} extended their work to two-way sparse reduced-rank regression. \citet{ChenHuang2016} extended the row-sparse reduced-rank regression by incorporating covariance matrix estimation, and the authors mainly focused on computational issues. \citet{LianFengZhao2015} proposed a semiparametric reduced-rank regression with a sparsity penalty on the coefficient matrix itself. \citet{GohChenDey2017} studied the Bayesian counterpart of the row/column-sparse reduced-rank regression and established its posterior consistency. However, none of these works considered the possible entrywise sparsity in the SVD of the coefficient matrix. The sparse and low-rank regression models have also been applied in various fields to solve important scientific problems. To name a few, \citet{ChenChanStenseth2014} applied a sparse and low-rank bi-linear model for the task of source-sink reconstruction in marine ecology, \citet{Zhuetal2014} used a Bayesian low-rank model for associating neuroimaging phenotypes and genetic markers, and \citet{MaXiaoWong2014} used a threshold SVD regression model for learning regulatory relationships in genomics.

In view of the key role that the sparse SVD plays for simultaneous dimension reduction and variable selection in model \eqref{eq:model}, in this paper we suggest a unified regularization approach to estimating such a sparse SVD structure. Our proposal successfully meets three key methodological challenges that are posed by the complex structural constraints on the SVD. First, sparsity and orthogonality are two largely incompatible goals and would seem difficult to be accommodated within a single framework. For instance, a standard orthogonalization process such as QR factorization will generally destroy the sparsity pattern of a matrix. Previous methods either relaxed the orthogonality constraint to allow efficient search for sparsity patterns \citep{Chen:Chan:Sten:redu:2012}, or avoided imposing both sparsity and orthogonality requirements on the same factor matrix \citep{Chen:Huan:spar:2012,Bune:She:Wegk:join:2012}. To resolve this issue, we formulate our approach as an orthogonality constrained regularization problem, which yields \textit{simultaneously sparse and orthogonal} factor matrices in the SVD. Second, we employ the nuclear 
norm penalty to encourage sparsity among the singular values and achieve rank reduction. As a result, our method produces a continuous solution path, which facilitates rank parameter tuning and distinguishes it from the $L_0$ regularization method adopted by \citet{Bune:She:Wegk:join:2012}. Third, unlike rank-constrained estimation, the nuclear norm penalization approach makes the estimation of singular vectors more intricate, since one does not know a priori which singular values will vanish and, hence, which pairs of left and right singular vectors are unidentifiable. Noting that the degree of identifiability of the singular vectors increases with the singular value, we propose to penalize the singular vectors weighted by singular values, which proves to be meaningful and effective. Combining these aspects, we introduce \textit{sparse orthogonal factor regression} (SOFAR), a novel regularization framework for high-dimensional multivariate regression. While respecting the orthogonality constraint, we allow the sparsity-inducing penalties to take a general, flexible form, which includes special cases that adapt to the entrywise and rowwise sparsity of the singular vector matrices, resulting in a nonconvex objective function for the SOFAR method.

In addition to the aforementioned three methodological challenges, the nonconvexity of the SOFAR objective function also poses important algorithmic and theoretical challenges in obtaining and characterizing the SOFAR estimator. To address these challenges, we suggest a two-step approach exploiting the framework of convexity-assisted nonconvex optimization (CANO) to obtain the SOFAR estimator. More specifically, in the first step we minimize the $L_1$-penalized squared loss for the multivariate regression  \eqref{eq:model} to obtain an initial estimator. Then in the second step, we minimize the SOFAR objective function in an asymptotically shrinking neighborhood of the initial estimator. Thanks to the convexity of its objective function, the initial estimator can be obtained effectively and efficiently. Yet since the finer sparsity structure imposed through the sparse SVD \eqref{eq:svd} is completely ignored in the first step,  the initial estimator meets none of the aforementioned three methodological challenges. Nevertheless, since it is theoretically guaranteed that the initial estimator is not far away from the true coefficient matrix $\bC^*$ with asymptotic probability one, searching in an asymptotically shrinking neighborhood of the initial estimator significantly alleviates the nonconvexity issue of the SOFAR objective function.
In fact, under the framework of CANO we derive nonasymptotic bounds for the prediction, estimation, and variable selection errors of the SOFAR estimator characterizing the theoretical advantages. {In implementation, to disentangle the sparsity and orthogonality constraints we develop an efficient SOFAR algorithm
and establish 
its convergence properties. 
}


Our suggested SOFAR method for large-scale association network learning is in fact connected to a variety of statistical methods in both unsupervised and supervised multivariate analysis. For example, the sparse SVD and sparse principal component analysis (PCA) for a high-dimensional data matrix 
can be viewed as unsupervised versions of our general method. Other prominent examples include sparse factor models, sparse canonical correlation analysis \citep{witten2009penalized},  and sparse vector autoregressive (VAR) models for high-dimensional time series. 
See Section \ref{sec:appl} for more details on these applications and  connections.

The rest of the paper is organized as follows. Section \ref{Sec2} introduces the SOFAR method and discusses its applications to several 
unsupervised and supervised learning tasks. We present the nonasymptotic properties of the method in Section \ref{Sec3}. Section \ref{Sec4} develops an efficient optimization algorithm and discusses its convergence and tuning parameter selection. We provide several simulation and real data examples in Section \ref{Sec5}. 
All the proofs of main results and technical details are detailed in the Supplementary Material. An associated R package implementing the suggested method is available at 
{\small \url{http://www-bcf.usc.edu/~jinchilv/publications/software}}.


\section{Large-scale association network learning via SOFAR} \label{Sec2}

\subsection{Sparse orthogonal factor regression} \label{Sec2.1}
To estimate the sparse SVD of the true regression coefficient matrix $\bC^*$ in model \eqref{eq:model}, we start by considering an estimator of the form $\bU\bD\bV^T$, where $\bD=\diag(d_1,\dots,d_m)\in\mathbb{R}^{m\times m}$ with $d_1\ge\dots\ge d_m\ge0$ and $1\leq m \leq \min\{p,q\}$ is a diagonal matrix of singular values, and $\bU=(\bu_1,\dots,\bu_m)\in\mathbb{R}^{p\times m}$ and $\bV=(\bv_1,\dots,\bv_m)\in\mathbb{R}^{q\times m}$ are orthonormal matrices of left and right singular vectors, respectively. Although it is always possible to take $m=\min(p,q)$ without prior knowledge of the rank $r$, it is often sufficient in practice to take a small $m$ that is slightly larger than the expected rank (estimated by some procedure such as in \citet{Bune:She:Wegk:opti:2011}), which can dramatically reduce computation time and space. Throughout the paper, for any matrix $\bM=(m_{ij})$ we denote by $\|\bM\|_F$, $\|\bM\|_1$, $\|\bM\|_{\infty}$, and $\|\bM\|_{2,1}$ the Frobenius norm, entrywise $L_1$-norm, entrywise $L_{\infty}$-norm, and rowwise $(2,1)$-norm defined, respectively, as  $\|\bM\|_F=\bigl(\sum_{i,j}m_{ij}^2\bigr)^{1/2}$, $\|\bM\|_1=\sum_{i,j}|m_{ij}|$, $\|\bM\|_{\infty}=\max_{i,j}|m_{ij}|$, and $\|\bM\|_{2,1}=\sum_i\bigl(\sum_j m_{ij}^2\bigr)^{1/2}$. We also denote by $\|\cdot\|_2$ the induced matrix norm (operator norm).

As mentioned in the Introduction, we employ the nuclear norm penalty to encourage sparsity among the singular values, which is exactly the entrywise $L_1$ penalty on $\bD$. Penalization directly on $\bU$ and $\bV$, however, is inappropriate since the singular vectors are not equally identifiable and should not be subject to the same amount of regularization. Singular vectors corresponding to larger singular values can be estimated more accurately and should contribute more to the regularization, whereas those corresponding to vanishing singular values are unidentifiable and should play no role in the regularization. Therefore, we propose an \textit{importance weighting} by the singular values and place sparsity-inducing penalties on the weighted versions of singular vector matrices, $\bU\bD$ and $\bV\bD$. Also taking into account the orthogonality constraints on $\bU$ and $\bV$, we consider the orthogonality constrained optimization problem
\begin{align}\label{eq:sofar}
\begin{split}
(\what\bD,\what\bU,\what\bV)&=\argmin_{\bD,\bU,\bV}\left\{\frac{1}{2}\|\bY-\bX\bU\bD\bV^T\|_F^2+\lambda_d\|\bD\|_1+\lambda_a\rho_a(\bU\bD) +\lambda_b\rho_b(\bV\bD)\right\}\\
&\relph{=}\text{subject to}\quad\bU^T\bU=\bI_m,\quad\bV^T\bV=\bI_m,
\end{split}
\end{align}
where $\rho_a(\cdot)$ and $\rho_b(\cdot)$ are penalty functions to be clarified later, and $\lambda_d,\lambda_a,\lambda_b\ge0$ are tuning parameters that control the strengths of regularization. We call this regularization method \textit{sparse orthogonal factor regression} (SOFAR) and the regularized estimator $(\what\bD,\what\bU,\what\bV)$ the SOFAR estimator. Note that $\rho_a(\cdot)$ and $\rho_b(\cdot)$ can be equal or distinct, depending on the scientific question and the goals of variable selection. Letting $\lambda_d=\lambda_b=0$ while setting $\rho_a(\cdot)=\|\cdot\|_{2,1}$ reduces the SOFAR estimator to the sparse reduced-rank estimator of \citet{Chen:Huan:spar:2012}. In view of our choices of $\rho_a(\cdot)$ and $\rho_b(\cdot)$, although $\bD$ appears in all three penalty terms, rank reduction is achieved mainly through the first term, while variable selection is achieved through the last two terms under necessary scalings by $\bD$.

Note that for simplicity we do not explicitly state the ordering constraint $d_1\ge\dots\ge d_m\ge0$ in optimization problem \eqref{eq:sofar}. In fact, when $\rho_a(\cdot)$ and $\rho_b(\cdot)$ are matrix norms that satisfy certain invariance properties, such as the entrywise $L_1$-norm and rowwise $(2,1)$-norm, this constraint can be easily enforced by simultaneously permuting and/or changing the signs of the singular values and the corresponding singular vectors. The orthogonality constraints are, however, essential to the optimization problem in that a solution cannot be simply obtained through solving the unconstrained regularization problem followed by an orthogonalization process. The interplay between sparse regularization and orthogonality constraints is crucial for achieving important theoretical and practical advantages, which distinguishes our SOFAR method from most previous procedures.

\subsection{Applications of SOFAR} \label{sec:appl}
The SOFAR method provides a unified framework for a variety of statistical problems in multivariate analysis. We give four such examples, and in each example, briefly review existing techniques and suggest new methods.

\subsubsection{Biclustering with sparse SVD} \label{Sec2.2.1}
The biclustering problem of a data matrix, which can be traced back to \citet{Hart:dire:1972}, aims to simultaneously cluster the rows (samples) and columns (features) of a data matrix into statistically related subgroups. A variety of biclustering techniques, which differ in the criteria used to relate clusters of samples and clusters of features and in whether overlapping of clusters is allowed, have been suggested as useful tools in the exploratory analysis of high-dimensional genomic and text data. See, for example, \citet{Busy:Prok:Pard:bicl:2008} for a survey. One way of formulating the biclustering problem is through the mean model
\begin{equation}\label{eq:mean}
\bX=\bC^*+\bE,
\end{equation}
where the mean matrix $\bC^*$ admits a sparse SVD \eqref{eq:svd} and the sparsity patterns in the left (or right) singular vectors serve as indicators for the samples (or features) to be clustered. \citet{Lee:Shen:Huan:Marr:bicl:2010} proposed to estimate the first sparse SVD layer by solving the optimization problem
\begin{align}\label{eq:lee}
\begin{split}
(\hat{d},\what\bu,\what\bv)&=\argmin_{d,\bu,\bv}\left\{\frac{1}{2}\|\bX-d\bu\bv^T\|_F^2+\lambda_a\rho_a(d\bu)+\lambda_b\rho_b(d\bv)\right\}\\
&\relph{=}\text{subject to}\quad\|\bu\|_2=1,\quad\|\bv\|_2=1,
\end{split}
\end{align}
and obtain the next sparse SVD layer by applying the same procedure to the residual matrix $\bX-\hat{d}\what\bu\what\bv^T$. Clearly, problem \eqref{eq:lee} is a specific example of the SOFAR problem \eqref{eq:sofar} with $m=1$ and $\lambda_d=0$; however, the orthogonality constraints are not maintained during the layer-by-layer extraction process. The orthogonality issue also exists in most previous proposals, for example, \citet{Zhan:Zha:Simo:low-:2002}.

The multivariate linear model \eqref{eq:model} with a sparse SVD \eqref{eq:svd} can be viewed as a supervised version of the above biclustering problem, which extends the mean model \eqref{eq:mean} to a general design matrix and can be used to identify interpretable clusters of predictors and clusters of responses that are significantly associated. Applying the SOFAR method to model \eqref{eq:mean} yields the new estimator
\begin{align}\label{eq:ident}
\begin{split}
(\what\bD,\what\bU,\what\bV)&=\argmin_{\bD,\bU,\bV}\left\{\frac{1}{2}\|\bX-\bU\bD\bV^T\|_F^2+\lambda_d\|\bD\|_1+\lambda_a\rho_a(\bU\bD) +\lambda_b\rho_b(\bV\bD)\right\}\\
&\relph{=}\text{subject to}\quad\bU^T\bU=\bI_m,\quad\bV^T\bV=\bI_m,
\end{split}
\end{align}
which estimates all sparse SVD layers simultaneously while determining the rank by nuclear norm penalization and preserving the orthogonality constraints.

\subsubsection{Sparse PCA} \label{Sec2.2.2}
A useful technique closely related to sparse SVD is sparse principal component analysis (PCA), which enhances the convergence and improves the interpretability of PCA by introducing sparsity in the loadings of principal components. There has been a fast growing literature on sparse PCA due to its importance in dimension reduction for high-dimensional data. Various formulations coupled with efficient algorithms, notably through $L_0$ regularization and its $L_1$ and semidefinite relaxations, have been proposed by \citet{Zou:Hast:Tibs:spar:2006}, \citet{Aspr:Ghao:Jord:Lanc:dire:2007}, \citet{Shen:Huan:spar:2008}, \citet{John:Lu:on:2009}, and \citet{Guo:Jame:Levi:Mich:Zhu:prin:2010}, among others.

We are interested in two different ways of casting sparse PCA in our sparse SVD framework. The first approach bears a resemblance to the proposal of \citet{Zou:Hast:Tibs:spar:2006}, which formulates sparse PCA as a regularized multivariate regression problem with the data matrix $\bX$ treated as both the responses and the predictors. Specifically, they proposed to solve the optimization problem
\begin{align}\label{eq:zou}
\begin{split}
(\what\bA,\what\bV)&=\argmin_{\bA,\bV}\left\{\frac{1}{2}\|\bX-\bX\bA\bV^T\|_F^2+\lambda_a\rho_a(\bA)\right\}\\
&\relph{=}\text{subject to}\quad\bV^T\bV=\bI_m,
\end{split}
\end{align}
and the loading vectors are given by the normalized columns of $\what\bA$, $\what\ba_j/\|\what\ba_j\|_2$, $j=1,\dots,m$. However, the orthogonality of the loading vectors, a desirable property enjoyed by the standard PCA, is not enforced by problem \eqref{eq:zou}. Similarly applying the SOFAR method leads to the estimator
\begin{align*}
(\what\bD,\what\bU,\what\bV)&=\argmin_{\bD,\bU,\bV}\left\{\frac{1}{2}\|\bX-\bX\bU\bD\bV^T\|_F^2+\lambda_d\|\bD\|_1+\lambda_a\rho_a(\bU\bD)\right\}\\
&\relph{=}\text{subject to}\quad\bU^T\bU=\bI_m,\quad\bV^T\bV=\bI_m,
\end{align*}
which explicitly imposes orthogonality among the loading vectors (the columns of $\what\bU$). One can optionally ignore the nuclear norm penalty and determine the number of principal components by some well-established criterion.

The second approach exploits the connection of sparse PCA with regularized SVD suggested by \citet{Shen:Huan:spar:2008}. They proposed to solve the rank-1 matrix approximation problem
\begin{align}\label{eq:shen}
\begin{split}
(\what\bu,\what\bb)&=\argmin_{\bu,\bb}\left\{\frac{1}{2}\|\bX-\bu\bb^T\|_F^2+\lambda_b\rho_b(\bb)\right\}\\
&\relph{=}\text{subject to}\quad\|\bu\|_2=1,
\end{split}
\end{align}
and obtain the first loading vector $\what\bb/\|\what\bb\|_2$. Applying the SOFAR method similarly to the rank-$m$ matrix approximation problem yields the estimator
\begin{align*}
(\what\bD,\what\bU,\what\bV)&=\argmin_{\bD,\bU,\bV}\left\{\frac{1}{2}\|\bX-\bU\bD\bV^T\|_F^2+\lambda_d\|\bD\|_1+\lambda_b\rho_b(\bV\bD)\right\}\\
&\relph{=}\text{subject to}\quad\bU^T\bU=\bI_m,\quad\bV^T\bV=\bI_m,
\end{align*}
which constitutes a multivariate generalization of problem \eqref{eq:shen}, with the desirable orthogonality constraint imposed on the loading vectors (the columns of $\what\bV$) and the optional nuclear norm penalty useful for determining the number of principal components.

\subsubsection{Sparse factor analysis} \label{Sec2.2.3}
Factor analysis plays an important role in dimension reduction and feature extraction for high-dimensional time series. A low-dimensional factor structure is appealing from both theoretical and practical angles, and can be conveniently incorporated into many other statistical tasks, such as forecasting with factor-augmented regression \citep{Stoc:Wats:fore:2002} and covariance matrix estimation \citep{Fan:Fan:Lv:high:2008}. See, for example, \citet{Bai:Ng:larg:2008} for an overview.

Let $\bx_t\in\mathbb{R}^p$ be a vector of observed time series. Consider the factor model
\begin{equation}\label{eq:factor}
\bx_t=\bLambda\bff_t+\be_t,\quad t=1,\dots,T,
\end{equation}
where $\bff_t\in\mathbb{R}^m$ is a vector of latent factors, $\bLambda\in\mathbb{R}^{p\times m}$ is the factor loading matrix, and $\be_t$ is the idiosyncratic error. Most existing methods for high-dimensional factor models rely on classical PCA \citep{Bai:Ng:2002,Bai:infe:2003} or maximum likelihood to estimate the factors and factor loadings \citep{Bai:Li:2016,Bai:Li:stat:2012}; as a result, the estimated factors and loadings are generally nonzero. However, in order to assign economic meanings to the factors and loadings and to further mitigate the curse of dimensionality, it would be desirable to introduce sparsity in the factors and loadings. Writing model \eqref{eq:factor} in the matrix form
\[
\bX=\bF\bLambda^T+\bE 
\]
with $\bX=(\bx_1,\dots,\bx_T)^T$, $\bF=(\bff_1,\dots,\bff_T)^T$, and $\bE=(\be_1,\dots,\be_T)^T$ reveals its equivalence to model \eqref{eq:mean}. Therefore, under the usual normalization restrictions that $\bF^T\bF/T=\bI_m$ and $\bLambda^T\bLambda$ is diagonal, we can solve for $(\what\bD,\what\bU,\what\bV)$ in problem \eqref{eq:ident} and estimate the sparse factors and loadings by $\what\bF=\sqrt{T}\what\bU$ and $\what\bLambda=\what\bV\what\bD/\sqrt{T}$.

\subsubsection{Sparse VAR analysis} \label{Sec2.2.4}
Vector autoregressive (VAR) models have been widely used to analyze the joint dynamics of multivariate time series; see, for example, \citet{Stoc:Wats:2001}. Classical VAR analysis suffers greatly from the large number of free parameters in a VAR model, which grows quadratically with the dimensionality. Early attempts in reducing the impact of dimensionality have explored reduced rank methods such as canonical analysis and reduced rank regression \citep{Box:Tiao:cano:1977,Velu:Rein:Wich:redu:1986}. Regularization methods such as the Lasso have recently been adapted to VAR analysis for variable selection \citep{Hsu:Hung:Chan:subs:2008,Nard:Rina:auto:2011,Kock:Callot:2015,Basu:Michailidis:2015}.

We present an example in which our parsimonious model setup is most appropriate. Suppose we observe the data $(\by_t,\bx_t)$, where $\by_t\in\mathbb{R}^q$ is a low-dimensional vector of time series whose dynamics are of primary interest, and $\bx_t\in\mathbb{R}^p$ is a high-dimensional vector of informational time series. We assume that $\bx_t$ are generated by the VAR equation
\[
\bx_t=\bC^{*T}\bx_{t-1}+\be_t,
\]
where $\bC$ has a sparse SVD \eqref{eq:svd}. This implies a low-dimensional latent model of the form
\[
\bg_t=\bD^*\bff_{t-1}+\wtilde\be_t,
\]
where $\bff_t=\bU^{*T}\bx_t$, $\bg_t=\bV^{*T}\bx_t$, and $\wtilde\be_t=\bV^{*T}\be_t$. Following the factor-augmented VAR (FAVAR) approach of \citet{Bern:Boiv:Elia:meas:2005}, we augment the latent factors $\bff_t$ and $\bg_t$ to the dynamic equation of $\by_t$ and consider the joint model
\[
\begin{pmatrix}
\by_t\\
\bg_t
\end{pmatrix}=\begin{pmatrix}
\bA^T & \bB^T\\
\bzero & \bD^*
\end{pmatrix}\begin{pmatrix}
\by_{t-1}\\
\bff_{t-1}
\end{pmatrix}+\begin{pmatrix}
\bve_t\\
\wtilde\be_t
\end{pmatrix}.
\]
We can estimate the parameters $\bA$, $\bB$, and $\bD^*$ by a two-step method: first apply the SOFAR method to obtain estimates of $\bD^*$ and $\bff_t$, and then estimate $\bA$ and $\bB$ by a usual VAR since both $\by_t$ and $\bff_t$ are of low dimensionality. Our approach differs from previous methods in that we enforce sparse factor loadings; hence, it would allow the factors to be given economic interpretations and would be useful for uncovering the structural relationships underlying the joint dynamics of $(\by_t,\bx_t)$.

\section{Theoretical properties}\label{sec:theory} \label{Sec3}
We now investigate the theoretical properties of the SOFAR estimator \eqref{eq:sofar} for model \eqref{eq:model} under the sparse SVD structure \eqref{eq:svd}. Our results concern nonasymptotic error bounds, where both response dimensionality $q$ and predictor dimensionality $p$ can diverge simultaneously with sample size $n$. The major theoretical challenges stem from the nonconvexity issues of our optimization problem which are prevalent in nonconvex statistical learning. 

\subsection{Technical conditions} \label{Sec3.1}
We begin with specifying a few assumptions that facilitate our technical analysis. To simplify the technical presentation, we focus on the scenario of $p\geq q$ and our proofs can be adapted easily to the case of $p < q$ with the only difference that the rates of convergence in Theorems \ref{thm:initial} and \ref{thm:det} will be modified correspondingly.  
Assume that each column of $\bX$, $\tbx_j$ with $j=1,\dots,p$, has been rescaled such that $\|\tbx_j\|_2^2=n$.
The SOFAR method minimizes the objective function in \eqref{eq:sofar}.   Since the true rank $r$ is unknown and we cannot expect that one can choose $m$ to perfectly match $r$, the SOFAR estimates $\what{\bU}$, $\what{\bV}$, and $\what{\bD}$ are generally of different sizes than $\bU^*$, $\bV^*$, and $\bD^*$, respectively. To ease the presentation, we expand the dimensions of matrices $\bU^*$, $\bV^*$, and $\bD^*$ by simply adding columns and rows of zeros to the right and to the bottom of each of the matrices to make them of sizes $p\times q$, $q\times q$, and $q\times q$, respectively.  We also expand the matrices $\what\bD$, $\what{\bU}$, and $\what{\bV}$ similarly to match the sizes of $\bD^*$, $\bU^*$, and $\bV^*$, respectively. Define $\bA^*=\bU^*\bD^*$ and $\bB^*=\bV^*\bD^*$, and correspondingly $\what{\bA} = \what{\bU}\what{\bD}$ and $\what{\bB} = \what{\bV}\what{\bD}$ using the SOFAR estimates $(\what{\bU}, \what{\bV}, \what{\bD})$. 

\begin{definition}[Robust spark] \label{def1}
	The robust spark $\kappa_c$ of the $n\times p$ design matrix $\bX$ is defined as the smallest possible positive integer such that there exists an $n\times \kappa_c$ submatrix of $n^{-1/2}\bX$ having a singular value less than a given positive constant
	$c$.
\end{definition}

\begin{condition}{\it(Parameter space)} \label{Aparsp}
The true parameters $(\bC^*,\bD^*,\bA^*,\bB^*)$ lie in $\mathcal{C} \times \mathcal{D} \times \mathcal{A} \times \mathcal{B}$, where 
$
\mathcal{C} = \{ \bC \in \mathbb{R}^{p\times q} : \|\bC\|_0 < \kappa_{c_2}/2\}$, 
$\mathcal{D} = \{ \bD = \diag\{d_j\} \in \mathbb{R}^{q\times q} : 
d_{j}=0 \mbox{ or } |d_{j}| \geq \tau 
\}$, 
$\mathcal{A} = \{ \bA = (a_{ij}) \in \mathbb{R}^{p\times q} : a_{ij}=0 \mbox{ or } |a_{ij}| \geq \tau 
\}$, and
$\mathcal{B} = \{ \bB = (b_{ij})\in \mathbb{R}^{q\times q} : b_{ij}=0 \mbox{ or } |a_{ij}| \geq \tau 
\}$ 
with $\kappa_{c_2}$ the robust spark  of $\bX$, $c_2 > 0$ some constant, and $\tau > 0$ asymptotically vanishing. 
\end{condition}

\begin{condition}{\it (Constrained eigenvalue)} \label{Aeigen}
It holds that 
$
\max_{\|\bu\|_0<\kappa_{c_2}/2,~\|\bu\|_2=1}\| \bX\bu\|_2^2\leq c_3n$ and $ \max_{1\leq j\leq r}\|\bX\bu_j^*\|_2^2 \leq c_3 n$ for some constant $c_3 > 0$, where  
$\bu_j^*$ is the left singular vector of $\bC^*$ corresponding to singular value $d_j^*$.
\end{condition}

\begin{condition}{\it(Error term)} \label{Aerror} The error term $\bE\in \mathbb{R}^{n\times q} \sim N(\bzero,\bI_n \otimes \bSigma)$ with the maximum eigenvalue $\alpha_{\max}$ of $\bSigma$ bounded from above and diagonal entries of $\bSigma$ being $\sigma_j^2$'s.
\end{condition}

\begin{condition}{\it(Penalty functions)} \label{Apen}
For matrices $\bM$ and $\bM^*$ of the same size,
the penalty functions $\rho_h$ with $h\in\{a,b\}$ satisfy $|\rho_h(\bM) - \rho_h(\bM^*)| \leq \|\bM-\bM^*\|_1$.
\end{condition}

\begin{condition}{\it(Relative spectral gap)} \label{Asgap}
The nonzero singular values of $\bC^*$ 
satisfy that
$
d_{j-1}^{*2}-d_{j}^{*2} \geq \delta^{1/2} d_{j-1}^{*2}
$
for $2 \leq j \leq r$ with $\delta > 0$ some constant, and $r$ and $\sum_{j=1}^r(d_1^*/d_j^*)^2$ can 
diverge as $n\rightarrow \infty$. 
\end{condition}

The concept of robust spark in Definition \ref{def1} was introduced initially  in \cite{Zheng:Fan:Lv:2014} and \cite{Fan:Lv:asym:2013}, where the thresholded parameter space was exploited to characterize the global optimum for regularization methods with general penalties.  Similarly, the thresholded parameter space and the constrained eigenvalue condition which builds on the robust spark condition of the design matrix in Conditions \ref{Aparsp} and \ref{Aeigen} are essential for investigating the 
computable solution to the nonconvex SOFAR optimization problem in \eqref{eq:sofar}. By Proposition 1 of \cite{Fan:Lv:asym:2013}, the robust spark $\kappa_{c_2}$ can be at least of order $O\{n/(\log p)\}$ with asymptotic probability one when the rows of $\bX$ are independently sampled from multivariate Gaussian distributions with dependency.
Although Condition \ref{Aerror} assumes Gaussianity, our theory can in principle carry over to the case of sub-Gaussian errors, provided that the concentration inequalities for Gaussian random variables used in our proofs are replaced by those for sub-Gaussian random variables.

Condition \ref{Apen} includes many kinds of penalty functions that bring about sparse estimates.
Important examples include the entrywise $L_1$-norm and rowwise $(2,1)$-norm, where the former encourages sparsity among the predictor/response effects specific to each rank-1 SVD layer, while the latter promotes predictor/response-wise sparsity regardless of the specific layer.
To see why the rowwise $(2,1)$-norm satisfies Condition \ref{Apen}, observe that
\begin{align*}
\|\bM\|_1 \equiv \sum_i \sum_j |m_{ij}| = \sum_i \left( \sum_{j,k} |m_{ij}||m_{ik}| \right)^{1/2} 
\geq \sum_i \left( \sum_jm_{ij}^2 \right)^{1/2}
\equiv \|\bM\|_{2,1},
\end{align*}
which along with the triangle inequality entails that Condition \ref{Apen} is indeed satisfied. Moreover, Condition \ref{Apen} allows us to use concave penalties such as SCAD \citep{FanLi2001} and MCP \citep{Zhang2010};  
see, for instance, the proof of Lemma 1 in \cite{Fan:Lv:asym:2013}.

Intuitively, Condition \ref{Asgap} rules out the nonidentifiable case where some nonzero singular values are tied with each other and the associated singular vectors in matrices $\bU^*$ and $\bV^*$ are identifiable only up to some orthogonal transformation. In particular, Condition \ref{Asgap} enables us to establish the key Lemma \ref{lem:MPT} in Section \ref{secB.3} of Supplementary Material, where the matrix perturbation theory can be invoked.

\subsection{Main results} \label{Sec3.2}

Since the objective function of the SOFAR method \eqref{eq:sofar} is nonconvex, solving this optimization problem is highly 
challenging. To overcome the difficulties, as mentioned in the Introduction we exploit the framework of CANO and suggest a two-step approach, where in the first step we solve the following $L_1$-penalized squared loss minimization problem 
\begin{align}\label{eq: l1-obj}
\wtilde{\bC} = \argmin_{\bC \in \mathbb{R}^{p\times q}} \left\{ (2n)^{-1} \| \bY - \bX \bC \|_F^2 + \lambda_0 \| \bC \|_1\right\}
\end{align}
to construct an initial estimator $\widetilde{\bC}$ with $\lambda_0 \geq 0$ some regularization parameter. If $\widetilde{\bC} = \bzero$, then we set the final SOFAR estimator as $\widehat{\bC} = \bzero$; otherwise, in the second step we do a refined search and minimize the SOFAR objective function \eqref{eq:sofar} in an asymptotically shrinking neighborhood of $\widetilde{\bC}$ to obtain the final SOFAR estimator $\widehat{\bC}$. In the case of $\widetilde{\bC} = \bzero$, our two-step procedure reduces to a one-step procedure. Since Theorem \ref{thm:initial} below establishes that  $\widetilde{\bC}$ can be close to $\bC^*$ with asymptotic probability one, having $\widetilde{\bC} = \bzero$ is a good indicator that the true $\bC^* = \bzero$.

Thanks to its convexity, the objective function in \eqref{eq: l1-obj} in the first step can be solved easily and efficiently. In fact, since the objective function in \eqref{eq: l1-obj} is separable it follows that the $j$th column of $\widetilde{\bC}$ can be obtained by solving the univariate response Lasso regression
\begin{align*}
\min_{\bbeta\in \mathbb{R}^p}\left\{(2n)^{-1} \|\bY \be_j - \bX\bbeta\|_2^2 + \lambda_0\|\bbeta\|_1 \right\},
\end{align*}
where $\be_j$ is a $q$-dimensional vector with $j$th component 1 and all other components 0. The above univariate response Lasso regression has been studied extensively and well understood, and many efficient algorithms have been proposed for solving it. Denote by $(\wtilde\bD,\wtilde\bU,\wtilde\bV)$ the initial estimator of $(\bD^*,\bU^*,\bV^*)$ obtained from the SVD of $\widetilde{\bC}$, and let $\wtilde\bA=\wtilde\bU\wtilde\bD$ and $\wtilde\bB=\wtilde\bV\wtilde\bD$. Since the bounds for the SVD are key to the analysis of SOFAR estimator in the second step, for completeness we present the nonasymptotic bounds on estimation errors of the initial estimator in the following theorem.

\begin{theorem}[Error bounds for initial estimator] \label{thm:initial}
Assume that Conditions \ref{Aparsp}--\ref{Aerror} hold and
let $\lambda_0 = c_0\sigma_{\max} \left(n^{-1}\log (pq) \right)^{1/2}$
with $\sigma_{\max}=\max_{1\leq j \leq q}\sigma_j$ and $c_0> \sqrt{2}$ some constant. Then with probability
at least $1- 2(pq)^{1-c_0^2/2}$, the estimation error is bounded as
\begin{align}\label{eq: Lasso-err}
\|\widetilde{\bC}-\bC^* \|_F
\leq R_n \equiv c(n^{-1} s\log (pq))^{1/2}
\end{align}
with $s = \|\bC^*\|_0$ and $c > 0$ some constant. Under addition Condition \ref{Asgap}, with the same probability bound the following estimation error bounds hold simultaneously
\begin{align}
& \|\widetilde{\bD} - \bD^*\|_F \leq c(n^{-1} s\log (pq))^{1/2}, \label{eq: ini-D-err}\\
& \|\widetilde{\bA} - \bA^*\|_F + \|\widetilde{\bB} - \bB^*\|_F
  \leq c\eta_n(n^{-1} s\log (pq))^{1/2}, \label{eq: ini-AB-err}
\end{align}
where $\eta_n =1+\delta^{-1/2}\big( \sum_{j=1}^r(d_1^*/d_j^*)^2\big)^{1/2}$.
\end{theorem}

For the case of $q=1$, the estimation error bound \eqref{eq: Lasso-err} is consistent with the well-known oracle inequality for Lasso \citep{Bickel:Ritov:Tsybakov:2009}. The additional estimation error bounds \eqref{eq: ini-D-err} and \eqref{eq: ini-AB-err} for the SVD in Theorem \ref{thm:initial} are, however, new to the literature. It is worth mentioning that Condition \ref{Asgap} and the latest results in \cite{Yu:Wang:Samworth:2015} play a crucial role in establishing these additional error bounds.


After obtaining the initial estimator $\wtilde{\bC}$ from the first step, we can solve the SOFAR optimization problem in an asymptotically shrinking  neighborhood of $\wtilde{\bC}$. More specifically, we define   $\wtilde{\mathcal{P}}_n = \{\bC: \|\bC - \wtilde\bC\|_F \leq 2R_n\}$ with  $R_n$ the upper bound in \eqref{eq: Lasso-err}. Then it is seen from Theorem \ref{thm:initial} that the true coefficient matrix $\bC^*$ is contained in $\wtilde{\mathcal{P}}_n$ with probability at least $1-2(pq)^{1-c_0^2/2}$.  Further define
\begin{equation}\label{def: local-neighbor}
\mathcal{P}_n = \wtilde{\mathcal{P}}_n \cap (\mathcal{C} \times \mathcal{D} \times \mathcal{A} \times \mathcal{B}),
\end{equation}
where sets $\mathcal{C}$, $\mathcal{D}$, $\mathcal{A}$, and $\mathcal{B}$ are defined in Condition \ref{Aparsp}. Then with probability at least $1-2(pq)^{1-c_0^2/2}$, the set $\mathcal{P}_n$ defined in (\ref{def: local-neighbor}) is nonempty  with at least one element $\bC^*$ by Condition \ref{Aparsp}. We minimize the SOFAR objective function \eqref{eq:sofar} by searching in the shrinking neighborhood $\mathcal{P}_n$ and denote by $\widehat{\bC}$ the resulting SOFAR estimator. Then it follows that with probability at least $1-2(pq)^{1-c_0^2/2}$,
\begin{align*}
 \|\widehat{\bC} - \bC^*\|_F \leq \|\widehat{\bC} - \widetilde{\bC}\|_F + \|\widetilde{\bC} - \bC^*\|_F \leq 3R_n,
\end{align*}
where the first inequality is by the triangle inequality and the second one is by the construction of set $\mathcal{P}_n$ and Theorem \ref{thm:initial}. Therefore, we see that the SOFAR estimator given by our two-step procedure is guaranteed to have convergence rate at least $O(R_n)$.

Since the initial estimator investigated in Theorem \ref{thm:initial} completely ignores the finer sparse SVD structure of the coefficient matrix $\bC^*$, intuitively the second step of SOFAR estimation can lead to improved error bounds. Indeed we show in Theorem \ref{thm:det} below that with the second step of refinement, up to some columnwise sign changes the SOFAR estimator can admit estimator error bounds in terms of parameters $r$, $s_a$, and $s_b$  with
$r=\|\bD^*\|_0$, $s_a=\|\bA^*\|_0$, and $s_b=\|\bB^*\|_0$. When $r$, $s_a$, and $s_b$ are drastically smaller than $s$, these new upper bounds can have better rates of convergence.

\begin{theorem}[Error bounds for SOFAR estimator]\label{thm:det}
Assume that Conditions \ref{Aparsp}--\ref{Asgap} hold, 
$
\lambda_{\max} \\\equiv \max(\lambda_d, \lambda_a, \lambda_b)  = c_1 \left( n^{-1}\log (pr)\right)^{1/2}$ with $c_1>0$ some large constant, $\log p= O(n^\alpha)$, $q= O(n^{\beta/2})$, $s= O(n^\gamma)$, and
$\eta_n^2 = o(\min\{\lambda_{\max}^{-1}\tau, n^{1-\alpha-\beta-\gamma}\tau^2\})$ 
with $\alpha, \beta, \gamma \geq 0$, $\alpha + \beta + \gamma < 1$, and $\eta_n$ as given in Theorem  \ref{thm:initial}.
Then with probability at least
\begin{align}\label{inequalities-probability}
& 1 -
\left\{ 2(pq)^{1-c_0^2/2} +
2(pr)^{-\tilde{c}_2}
+ 2pr \exp\left( -\tilde{c}_3n^{1-\beta - \gamma}\tau^{2}\eta_n^{-2}
\right)
\right\}, 
\end{align}
the SOFAR estimator satisfies the following error bounds simultaneously:
\begin{align}
(a)~~~& \|\what{\bC} - \bC^*\|_F \leq c\min\{s, (r + s_a+s_b)\eta_n^2\}^{1/2}\{n^{-1}\log(pq)\}^{1/2}, \label{eq: upperbound-1}\\
\nonumber (b)~~~&\|\what{\bD}-\bD^*\|_F +\|\what{\bA}-\bA^*\|_F+\|\what{\bB}-\bB^*\|_F\\
& \qquad \leq c\min\{s, (r + s_a+s_b)\eta_n^2\}^{1/2}\eta_n\{n^{-1}\log(pq)\}^{1/2},  \label{eq: upperbound-2} \\
(c)~~~& \|\what{\bD}-\bD^*\|_0 +\|\what{\bA}-\bA^*\|_0+\|\what{\bB}-\bB^*\|_0 \leq c(r+s_a+s_b), \label{eq: upperbound-3}\\
(d)~~~&\|\what{\bD}-\bD^*\|_1 +\|\what{\bA}-\bA^*\|_1+\|\what{\bB}-\bB^*\|_1  \leq c(r+s_a+s_b)\eta_n^2\lambda_{\max},\\
(e)~~~&n^{-1}\|\bX(\what{\bC}-\bC^*)\|_F^2  \leq c(r+s_a+s_b)\eta_n^2\lambda_{\max}^2, \label{eq: upperbound-4}
\end{align}
where $c_0>\sqrt{2}$ and $c, \tilde{c}_2, \tilde{c}_3$ are some positive  constants.
\end{theorem}

We see from Theorem \ref{thm:det} that the upper bounds in \eqref{eq: upperbound-1} and \eqref{eq: upperbound-2} are the minimum of two rates, one involving  $r + s_a + s_b$ (the total sparsity of $\bD^*$, $\bA^*$, and $\bB^*$) and the other one involving $s$ (the sparsity of matrix $\bC^*$). The rate involving $s$ is from the first step of Lasso estimation, while the rate involving $r+s_a + s_b$ is from the second step of SOFAR refinement. For the case of $s > (r+s_a + s_b)\eta_n^2$, our two-step procedure leads to enhanced error rates under the Frobenius norm. Moreover, the error rates in \eqref{eq: upperbound-3}--\eqref{eq: upperbound-4} are new to the literature and not shared by the initial Lasso estimator, showing again the advantages of having the second step of refinement. It is seen that our two-step SOFAR estimator is capable of recovering the sparsity structure of $\bD^*$, $\bA^*$, and $\bB^*$ very well.

Let us gain more insights into these new error bounds. In the case of univariate response with $q=1$, we have $\eta_n = 1 + \delta$, $r=1$, $s_a = s$, and $s_b = 1$. Then the upper bounds in \eqref{eq: upperbound-1}--\eqref{eq: upperbound-4} reduce to $c\{sn^{-1}\log p\}^{1/2}$, $c\{sn^{-1}\log p\}^{1/2}$, $cs$, $cs\{n^{-1}\log p\}^{1/2}$, and $cn^{-1}s\log p$,  respectively, which are indeed within a logarithmic factor of the oracle rates for the case of high-dimensional univariate response regression. Furthermore, in the rank-one case of $r = 1$ we have $\eta_n = 1 + \delta^{-1/2}$ and $s = s_as_b$. Correspondingly, the upper bounds in \eqref{eq: Lasso-err}--\eqref{eq: ini-AB-err}  for the initial Lasso estimator all become $c\{n^{-1}s_as_b\log(pq)\}^{1/2}$, while the upper bounds in \eqref{eq: upperbound-1}--\eqref{eq: upperbound-4} for the SOFAR estimator become $c\{(s_a+s_b)n^{-1}\log (pq)\}^{1/2}$, $c\{(s_a+s_b)n^{-1}\log (pq)\}^{1/2}$, $c(s_a+s_b)$, $c(s_a + s_b)\{n^{-1}\log (pq)\}^{1/2}$, and $cn^{-1}(s_a + s_b)\log (pq)$,  respectively. In particular, we see that the SOFAR estimator can have much improved rates of convergence even in the setting of $r=1$.

\section{Implementation of SOFAR} \label{Sec4}
The interplay between sparse regularization and orthogonality constraints creates substantial algorithmic challenges for solving the SOFAR optimization problem \eqref{eq:sofar}, for which many existing algorithms can become either inefficient or inapplicable. For example, coordinate descent methods that are popular for solving large-scale sparse regularization problems \citep{Frie:Hast:Hfli:Tibs:path:2007} are not directly applicable because the penalty terms in problem \eqref{eq:sofar} are not separable under the orthogonality constraints. Also, the general framework for algorithms involving orthogonality constraints \citep{Edel:Aria:Smit:geom:1998} does not take sparsity into account and hence does not lead to efficient algorithms in our context. Inspired by a recently revived interest in the augmented Lagrangian method (ALM) and its variants for large-scale optimization in statistics and machine learning \citep{Boyd:Pari:Chu:Pele:Ecks:dist:2011}, in this section we develop an efficient algorithm for solving problem \eqref{eq:sofar}. 

\subsection{SOFAR algorithm with ALM-BCD} \label{Sec4.1}
The architecture of the proposed SOFAR algorithm is based on the ALM coupled with block coordinate descent (BCD). The first construction step is to utilize variable splitting to separate the orthogonality constraints and sparsity-inducing penalties into different subproblems, which then enables efficient optimization in a block coordinate descent fashion. To this end, we introduce two new variables $\bA$ and $\bB$, and express problem \eqref{eq:sofar} in the equivalent form
\begin{align}\label{eq:sofar1}
\begin{split}
(\what\bTheta,\what\bOmega)&=\argmin_{\bTheta,\bOmega}\left\{\frac{1}{2}\|\bY-\bX\bU\bD\bV^T\|_F^2+\lambda_d\|\bD\|_1+\lambda_a\rho_a(\bA) +\lambda_b\rho_b(\bB)\right\}\\
&\relph{=}\text{subject to}\quad\bU^T\bU=\bI_m,\quad\bV^T\bV=\bI_m,\quad\bU\bD=\bA,\quad\bV\bD=\bB,
\end{split}
\end{align}
where $\bTheta=(\bD,\bU,\bV)$ and $\bOmega=(\bA,\bB)$. We form the augmented Lagrangian for problem \eqref{eq:sofar1} as
\begin{align*}
L_{\mu}(\bTheta,\bOmega,\bGamma)&=\frac{1}{2}\|\bY-\bX\bU\bD\bV^T\|_F^2+\lambda_d\|\bD\|_1+\lambda_a\rho_a(\bA)+\lambda_b\rho_b(\bB) +\langle\bGamma_a,\bU\bD-\bA\rangle\\
&\relph{=}{}+\langle\bGamma_b,\bV\bD-\bB\rangle+\frac{\mu}{2}\|\bU\bD-\bA\|_F^2+\frac{\mu}{2}\|\bV\bD-\bB\|_F^2,
\end{align*}
where $\bGamma=(\bGamma_a,\bGamma_b)$ is the set of Lagrangian multipliers and $\mu>0$ is a penalty parameter. Based on ALM, the proposed algorithm consists of the following iterations:
\begin{enumerate}
\item $(\bTheta,\bOmega)$-step: $(\bTheta^{k+1},\bOmega^{k+1})\gets\argmin_{\bTheta\colon\bU^T\bU=\bV^T\bV=\bI_m,\bOmega}L_{\mu}(\bTheta,\bOmega,\bGamma^k)$;
\item $\bGamma$-step: $\bGamma_a^{k+1}\gets\bGamma_a^k+\mu(\bU^{k+1}\bD^{k+1}-\bA^{k+1})$ and $\bGamma_b^{k+1}\gets\bGamma_b^k+\mu(\bV^{k+1}\bD^{k+1}-\bB^{k+1})$.
\end{enumerate}
The $(\bTheta,\bOmega)$-step can be solved by a block coordinate descent method \citep{Tsen:conv:2001} cycling through the blocks $\bU$, $\bV$, $\bD$, $\bA$, and $\bB$. Note that the orthogonality constraints and the sparsity-inducing penalties are now separated into subproblems with respect to $\bTheta$ and $\bOmega$, respectively. To achieve convergence of the SOFAR algorithm in practice, an inexact minimization with a few block coordinate descent iterations is often sufficient. Moreover, to enhance the convergence of the algorithm to a feasible solution we optionally increase the penalty parameter $\mu$ by a ratio $\gamma>1$ at the end of each iteration. This leads to the SOFAR algorithm with ALM-BCD described in Table \ref{alg:admm}.


\begin{table}
\hskip-3pt\caption{\label{alg:admm} SOFAR algorithm with ALM-BCD}
\fbox{%
\begin{tabular}{l}
\textit{Parameters}: $\lambda_d$, $\lambda_a$, $\lambda_b$, and $\gamma>1$\\
Initialize $\bU^0$, $\bV^0$, $\bD^0$, $\bA^0$, $\bB^0$, $\bGamma_a^0$, $\bGamma_b^0$, and $\mu^0$\\
For $k=0,1,\dots$ do\\
\quad update $\bU$, $\bV$, $\bD$, $\bA$, and $\bB$:\\
\qquad(a) $\bU^{k+1}\gets\argmin\limits_{\bU^T\bU=\bI_m}\left\{\frac{1}{2}\|\bY-\bX\bU\bD^k(\bV^k)^T\|_F^2 +\frac{\mu^k}{2}\|\bU\bD^k-\bA^k+\bGamma_a^k/\mu^k\|_F^2\right\}$\\
\qquad(b) $\bV^{k+1}\gets\argmin\limits_{\bV^T\bV=\bI_m}\left\{\frac{1}{2}\|\bY-\bX\bU^{k+1}\bD^k\bV^T\|_F^2 +\frac{\mu^k}{2}\|\bV\bD^k-\bB^k+\bGamma_b^k/\mu^k\|_F^2\right\}$\\
\qquad(c) $\bD^{k+1}\gets\argmin\limits_{\bD\ge\bzero}\Bigl\{\frac{1}{2}\|\bY-\bX\bU^{k+1}\bD(\bV^{k+1})^T\|_F^2 +\frac{\mu^k}{2}\|\bU^{k+1}\bD-\bA^k+\bGamma_a^k/\mu^k\|_F^2$\\
\qquad $\phantom{\text{(c) }\bD^{k+1}\gets\argmin\limits_{\bD\ge\bzero}\Bigl(}+\frac{\mu^k}{2}\|\bV^{k+1}\bD-\bB^k+\bGamma_b^k/\mu^k\|_F^2 +\lambda_d\|\bD\|_1\Bigr\}$\\
\qquad(d) $\bA^{k+1}\gets\argmin\limits_{\bA}\left\{\frac{\mu^k}{2}\|\bU^{k+1}\bD^{k+1}-\bA+\bGamma_a^k/\mu^k\|_F^2+\lambda_a\rho_a(\bA)\right\}$\\
\qquad(e) $\bB^{k+1}\gets\argmin\limits_{\bB}\left\{\frac{\mu^k}{2}\|\bV^{k+1}\bD^{k+1}-\bB+\bGamma_b^k/\mu^k\|_F^2+\lambda_b\rho_b(\bB)\right\}$\\
\qquad(f) optionally, repeat (a)--(e) until convergence\\
\quad update $\bGamma_a$ and $\bGamma_b$:\\
\qquad(a) $\bGamma_a^{k+1}\gets\bGamma_a^k+\mu^k(\bU^{k+1}\bD^{k+1}-\bA^{k+1})$\\
\qquad(b) $\bGamma_b^{k+1}\gets\bGamma_b^k+\mu^k(\bV^{k+1}\bD^{k+1}-\bB^{k+1})$\\
\quad update $\mu$ by $\mu^{k+1}\gets\gamma\mu^k$\\
end
\end{tabular}}
\end{table}

We still need to solve the subproblems in algorithm \ref{alg:admm}. The $\bU$-update is similar to the weighted orthogonal Procrustes problem considered by \citet{Kosc:Sway:weig:1991}. By expanding the squares and omitting terms not involving $\bU$, this subproblem is equivalent to minimizing
\[
\frac{1}{2}\|\bX\bU\bD^k\|_F^2-\tr(\bU^T\bX^T\bY\bV^k\bD^k)-\tr(\bU^T(\mu^k\bA^k-\bGamma_a^k)\bD^k)
\]
subject to $\bU^T\bU=\bI_m$. Taking a matrix $\bZ$ such that $\bZ^T\bZ=\rho^2\bI_p-\bX^T\bX$, where $\rho^2$ is the largest eigenvalue of $\bX^T\bX$, we can follow the argument of \citet{Kosc:Sway:weig:1991} to obtain the iterative algorithm: for $j=0,1,\dots$, form the $p\times m$ matrix $\bC_1=(\bX^T\bY\bV^k+\mu^k\bA^k-\bGamma_a^k+\bZ^T\bZ\bU^j\bD^k)\bD^k$, compute the SVD $\bU_1\bSigma_1\bV_1^T=\bC_1$, and update $\bU^{j+1}=\bU_1\bV_1^T$. Note that $\bC_1$ depends on $\bZ^T\bZ$ only, and hence the explicit computation of $\bZ$ is not needed. The $\bV$-update is similar to a standard orthogonal Procrustes problem and amounts to maximizing
\[
\tr(\bV^T\bY^T\bX\bU^{k+1}\bD^k)+\tr(\bV^T(\mu^k\bB^k-\bGamma_b^k)\bD^k)
\]
subject to $\bV^T\bV=\bI_m$. A direct method for this problem \citep[][pp.~327--328]{Golu:Van:matr:2013} gives the algorithm: form the $q\times m$ matrix $\bC_2=(\bY^T\bX\bU^{k+1}+\mu^k\bB^k-\bGamma_b^k)\bD^k$, compute the SVD $\bU_2\bSigma_2\bV_2^T=\bC_2$, and set $\bV=\bU_2\bV_2^T$. Since $m$ is usually small, the SVD computations in the $\bU$- and $\bV$-updates are cheap. The Lasso problem in the $\bD$-update reduces to a standard quadratic program with the nonnegativity constraint, which can be readily solved by efficient algorithms; see, for example, \citet{Sha:Lin:Saul:Lee:mult:2007}. Note that the $\bD$-update may set some singular values to exactly zero; hence, a greedy strategy can be taken to further bring down the computational complexity, by removing the zero singular values and reducing the sizes of the relevant matrices accordingly in subsequent computations. The updates of $\bA$ and $\bB$ are free of orthogonality constraints and therefore easy to solve. With the popular choices of $\|\cdot\|_1$ and $\|\cdot\|_{2,1}$ as the penalty functions, the updates can be performed by entrywise and rowwise soft-thresholding, respectively.

Following the theoretical analysis for the SOFAR method in Section \ref{sec:theory}, we employ the SVD of the cross-validated $L_1$-penalized estimator $\wtilde{\bC}$ in \eqref{eq: l1-obj} to initialize $\bU$, $\bV$, $\bD$, $\bA$, and $\bB$; the $\bGamma_a$ and $\bGamma_b$ are initialized as zero matrices. In practice, for large-scale problems we can further scale up the SOFAR method by performing feature screening with the initial estimator  $\wtilde{\bC}$, that is, the response variables corresponding to zero columns in $\wtilde{\bC}$ and the predictors corresponding to zero rows in $\wtilde{\bC}$ could be removed prior to the finer SOFAR analysis.

\subsection{Convergence analysis and tuning parameter selection} \label{Sec4.2}
For general nonconvex problems, an ALM algorithm needs not to converge, and even if it converges, it needs not to converge to an optimal solution. 
We have the following convergence results regarding the proposed SOFAR algorithm with ALM-BCD.

\begin{theorem}[Convergence of SOFAR algorithm]\label{thm:conv}
Assume that $\sum_{k = 1}^\infty \{[\Delta L_\mu(\bU^k)]^{1/2} + $\\ $ [\Delta L_\mu(\bV^k)]^{1/2} + [\Delta L_\mu(\bD^k)]^{1/2}\} < \infty$ and the penalty functions $\rho_a(\cdot)$ and $\rho_b(\cdot)$ are convex, where $\Delta L_\mu(\cdot)$ denotes the decrease in $L_\mu(\cdot)$ by a block update. Then the sequence generated by the SOFAR algorithm converges to a local solution of the augmented Lagrangian for problem \eqref{eq:sofar1}.
\end{theorem}

Note that without the above assumption on $(\bU^k)$, $(\bV^k)$, and $(\bD^k)$, we can only show that the differences between two consecutive $\bU$-, $\bV$-, and $\bD$-updates converge to zero by the convergence of the sequence $(L_\mu(\cdot))$, but the sequences $(\bU^k)$, $(\bV^k)$, and $(\bD^k)$ may not necessarily converge. Although Theorem \ref{thm:conv} does not ensure the convergence of algorithm \ref{alg:admm} to an optimal solution, numerical evidence suggests that the algorithm has strong convergence properties and the produced solutions perform well in numerical studies.



The above SOFAR algorithm is presented for a fixed triple of tuning parameters $(\lambda_d,\lambda_a,\lambda_b)$. One may apply a fine grid search with $K$-fold cross-validation or an information criterion such as BIC and its high-dimensional extensions including GIC \citep{Fan:Tang:tuni:2013} to choose an optimal triple of tuning parameters and hence a best model. In either case, a full search over a three-dimensional grid would be prohibitively expensive, especially for large-scale problems. Theorem \ref{thm:det}, however, suggests that the parameter tuning can be effectively reduced to one or two dimensions. Hence, we adopt a search strategy which is computationally affordable and still provides reasonable and robust performance. To this end, we first estimate an upper bound on each of the tuning parameters by considering the \textit{marginal null model}, where two of the three tuning parameters are fixed at zero and the other is set to the minimum value leading to a null model. We denote the upper bounds thus obtained by $(\lambda_d^*,\lambda_a^*,\lambda_b^*)$, and conduct a search over a one-dimensional grid of values between $(\lambda_d^*,\lambda_a^*,\lambda_b^*)$ and $(\ve\lambda_d^*,\ve\lambda_a^*,\ve\lambda_b^*)$, with $\ve>0$ sufficiently small (e.g., $10^{-3}$) to ensure the coverage of a full spectrum of reasonable solutions. Our numerical experience suggests that this simple search strategy works well in practice while reducing the computational cost dramatically. More flexibility can be gained by adjusting the ratios between $\lambda_d$, $\lambda_a$, and $\lambda_b$ if additional information about the relative sparsity levels of $\bD$, $\bA$, and $\bB$ is available.

\section{Numerical studies} \label{Sec5}

\subsection{Simulation examples} \label{Sec5.1.new}

Our Condition \ref{Apen} in Section \ref{Sec3.1} accommodates a large group of penalty functions including concave ones such as SCAD and MCP. 
As demonstrated in \cite{Zou:Li:one-:2008} and \cite{fan2014adaptive}, nonconvex regularization problems can be solved using the idea of local linear approximation, which essentially reduces the original problem to the weighted $L_1$-regularization with the weights chosen adaptively based on some initial solution.
For this reason, in the simulation study we focus on the entrywise $L_1$-norm $\|\cdot\|_1$ and the rowwise $(2,1)$-norm $\|\cdot\|_{2,1}$, as well as their adaptive extensions. 
The use of adaptively weighted penalties has also been explored in the contexts of reduced rank regression \citep{Chen:Dong:Chan:redu:2013} and sparse PCA \citep{Leng:Wang:on:2009}. We next provide more details on the adaptive penalties used in our simulation study. To simplify the presentation, we use the entrywise $L_1$-norm as an example.

 Incorporating adaptive weighting into the penalty terms in problem \eqref{eq:sofar1} leads to the adaptive SOFAR estimator
\begin{align*}
	(\what\bTheta,\what\bOmega)&=\argmin_{\bTheta,\bOmega}\left\{\frac{1}{2}\|\bY-\bX\bU\bD\bV^T\|_F^2+\lambda_d\|\bW_d\circ\bD\|_1+\lambda_a\|\bW_a\circ\bA\|_1 +\lambda_b\|\bW_b\circ\bB\|_1\right\}\\
	&\relph{=}\text{subject to}\quad\bU^T\bU=\bI_m,\quad\bV^T\bV=\bI_m,\quad\bU\bD=\bA,\quad\bV\bD=\bB,
\end{align*}
where $\bW_d\in\mathbb{R}^{m\times m}$, $\bW_a\in\mathbb{R}^{p\times m}$, and $\bW_b\in\mathbb{R}^{q\times m}$ are weighting matrices that depend on the initial estimates $\wtilde\bD$, $\wtilde\bA$, and $\wtilde\bB$, respectively, and $\circ$ is the Hadamard or entrywise product. The weighting matrices are chosen to reflect the intuition that singular values and singular vectors of larger magnitude should be less penalized in order to reduce bias and improve efficiency in estimation. As suggested in \cite{Zou:Li:one-:2008}, if one is interested in using some nonconvex penalty functions $\rho_a(\cdot)$ and $\rho_b(\cdot)$ then the weight matrices can be constructed by using the first order derivatives of the penalty functions and the initial solution $(\widetilde{\bA}, \widetilde{\bB}, \widetilde{\bD})$.
In our implementation, for simplification we adopt the alternative popular choice of $\bW_d=\diag(\widetilde{d}_1^{-1},\dots,\widetilde{d}_m^{-1})$ with  $\widetilde{d}_j$ the $j$th diagonal entry of $\wtilde\bD$, as suggested in \citet{Zou:adap:2006}. Similarly,  we set $\bW_a=(\widetilde{a}_{ij}^{-1})$ and $\bW_b=(\widetilde{b}_{ij}^{-1})$ with $\widetilde{a}_{ij}$ and $\widetilde{b}_{ij}$ the $(i,j)$th entries of $\wtilde\bA$ and $\wtilde\bB$, respectively. 
Extension of the SOFAR algorithm with ALM-BCD in Section \ref{Sec4.1} is also straightforward, with the $\bD$-update becoming an adaptive Lasso problem and the updates of $\bA$ and $\bB$ now performed by adaptive soft-thresholding. A further way of improving the estimation efficiency is to exploit regularization methods in the thresholded parameter space \citep{Fan:Lv:asym:2013} or thresholded regression \citep{Zheng:Fan:Lv:2014}, which we do not pursue in this paper.

We compare the SOFAR estimator with the entrywise $L_1$-norm (Lasso) penalty (SOFAR-L) or the rowwise $(2,1)$-norm (group Lasso) penalty (SOFAR-GL) with five alternative methods, including three classical methods, namely, the ordinary least squares (OLS), separate adaptive Lasso regressions (Lasso), and reduced rank regression (RRR), and two recent sparse and low rank methods, namely, reduced rank regression with sparse SVD (RSSVD) proposed by \citet{Chen:Chan:Sten:redu:2012} and sparse reduced rank regression (SRRR) considered by \citet{Chen:Huan:spar:2012} (see also the rank constrained group Lasso estimator in \citealt{Bune:She:Wegk:join:2012}). Both \citet{Chen:Chan:Sten:redu:2012} and \citet{Chen:Huan:spar:2012} used adaptive weighted penalization. We thus consider both nonadaptive and adaptive versions of the SOFAR-L, SOFAR-GL, RSSVD, and SRRR methods.

\subsubsection{Simulation setups} \label{Sec5.1}
We consider several simulation settings with various model dimensions and sparse SVD patterns in the coefficient matrix $\bC^*$. In all settings, we took the sample size $n=200$ and the true rank $r=3$. Models 1 and 2 concern the entrywise sparse SVD structure in $\bC^*$. The design matrix $\bX$ was generated with i.i.d.\ rows from $N_p(\bzero,\bSigma_x)$, where $\bSigma_x=(0.5^{|i-j|})$. In model 1, we set $p=100$ and $q=40$, and let $\bC^*=\sum_{j=1}^3d_j^*\bu_j^*\bv_j^{*T}$ with $d_1^*=20$, $d_2^*=15$, $d_3^*=10$, and
\begin{gather*}
\tilde\bu_1=(\unif(S_u,5),\rep(0,20))^T,\quad\tilde\bu_2=(\rep(0,3),-\tilde{u}_{1,4},\tilde{u}_{1,5},\unif(S_u,3),\rep(0,17))^T,\\
\tilde\bu_3=(\rep(0,8),\unif(S_u,2),\rep(0,15))^T,\quad\bu_j^*=\tilde\bu_j/\|\tilde\bu_j\|_2,\quad j=1,2,3,\\
\tilde\bv_1=(\unif(S_v,5),\rep(0,10))^T,\quad\tilde\bv_2=(\rep(0,5),\unif(S_v,5),\rep(0,5))^T,\\
\tilde\bv_3=(\rep(0,10),\unif(S_v,5))^T,\quad\bv_j^*=\tilde\bv_j/\|\tilde\bv_j\|_2,\quad j=1,2,3,
\end{gather*}
where $\unif(S,k)$ denotes a $k$-vector with i.i.d.\ entries from the uniform distribution on the set $S$, $S_u=\{-1,1\}$, $S_v=[-1,-0.5]\cup[0.5,1]$, $\rep(\alpha,k)$ denotes a $k$-vector replicating the value $\alpha$, and $\tilde{u}_{j,k}$ is the $k$th entry of $\tilde\bu_j$. Model 2 is similar to Model 1 except with higher model dimensions, where we set $p=400$, $q=120$, and appended 300 and 80 zeros to each $\bu_j^*$ and $\bv_j^*$ defined above, respectively.

Models 3 and 4 pertain to the rowwise/columnwise sparse SVD structure in $\bC^*$. Also, we intend to study the case of approximate low-rankness/sparsity, by not requiring the signals be bounded away from zero. We generated $\bX$ with i.i.d.\ rows from $N_p(\bzero,\bSigma_x)$, where $\bSigma_x$ has diagonal entries $1$ and off-diagonal entries $0.5$. The rowwise sparsity patterns were generated in a similar way to the setup in \citet{Chen:Huan:spar:2012} except that we allow also the matrix of right singular vectors to be rowwise sparse, so that response selection may also be necessary. Specifically, we let $\bC^*=\bC_1\bC_2^T$, where $\bC_1\in\mathbb{R}^{p\times r}$ with i.i.d.\ entries in its first $p_0$ rows from $N(0,1)$ and the rest set to zero, and $\bC_2\in\mathbb{R}^{q\times r}$ with i.i.d.\ entries in its first $q_0$ rows from $N(0,1)$ and the rest set to zero.
We set $p=100$, $p_0=10$, $q=q_0=10$ in Model 3, and $p=400$, $p_0=10$, $q=200$, and $q_0=10$ in Model 4.

Finally, in all four settings, we generated the data $\bY$ from the model $\bY=\bX\bC^*+\bE$, where the error matrix $\bE$ has i.i.d.\ rows from $N_q(\bzero,\sigma^2\bSigma)$ with $\bSigma=(0.5^{|i-j|})$. In each simulation, $\sigma^2$ is computed to control the signal to noise ratio, defined as $\|d_r^*\bX\bu_r^*\bv_r^{*T}\|_F/\|\bE\|_F$, to be exactly $1$. The simulation was replicated 300 times in each setting.

All methods under comparison except OLS require selection of tuning parameters, which include the rank parameter in RRR, RSSVD, and SRRR and the regularization parameters in SOFAR-L, SOFAR-GL, RSSVD, and SRRR. To reveal the full potential of each method, we chose the tuning parameters based on the predictive accuracy evaluated on a large, independently generated validation set of size 2000. The results with tuning parameters chosen by cross-validation or GIC \citep{Fan:Tang:tuni:2013} were similar to those based on a large validation set, and hence are not reported.


The model accuracy of each method is measured by the mean squared error $\|\what\bC-\bC^*\|_F^2/(pq)$ for estimation (MSE-Est) and $\|\bX(\what\bC-\bC^*)\|_F^2/(nq)$ for prediction (MSE-Pred). The variable selection performance is characterized by the false positive rate (FPR\%) and false negative rate (FNR\%) in recovering the sparsity patterns of the SVD, that is, $\mathrm{FPR}=\mathrm{FP}/(\mathrm{TN}+\mathrm{FP})$ and $\mathrm{FNR}=\mathrm{FN}/(\mathrm{TP}+\mathrm{FN})$, where TP, FP, TN, and FN are the numbers of true nonzeros, false nonzeros, true zeros, and false zeros, respectively. The rank selection performance is evaluated by average estimated rank (Rank) and the percentage of correct rank identification (Rank\%). Finally, for the SOFAR-L, SOFAR-GL, and RSSVD methods which explicitly produce an SVD, the orthogonality of estimated factor matrices is measured by $100(\|\what\bU^T\what\bU\|_1+\|\what\bV^T\what\bV\|_1-2r)$ (Orth), which is minimized at zero when exact orthogonality is achieved.

\subsubsection{Simulation results} \label{Sec5.2}

\begin{figure}
	\includegraphics[width=.5\textwidth]{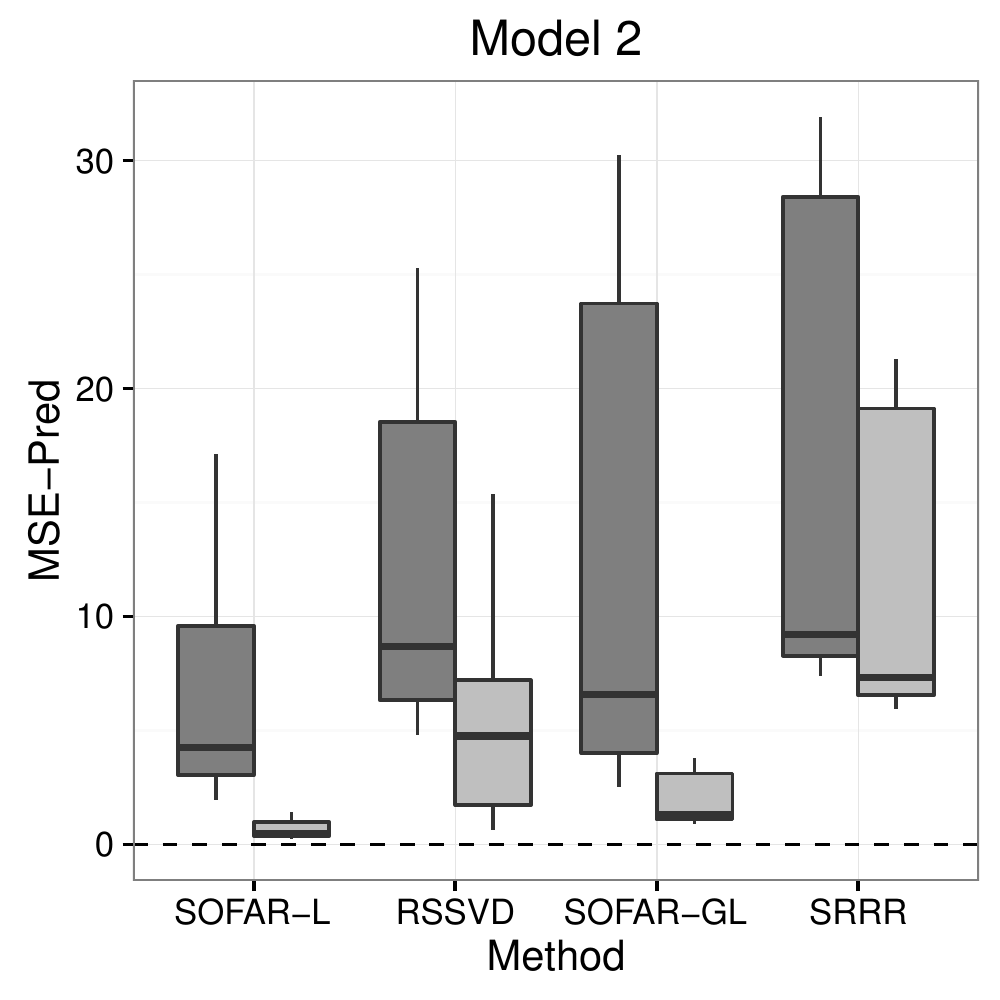}\includegraphics[width=.5\textwidth]{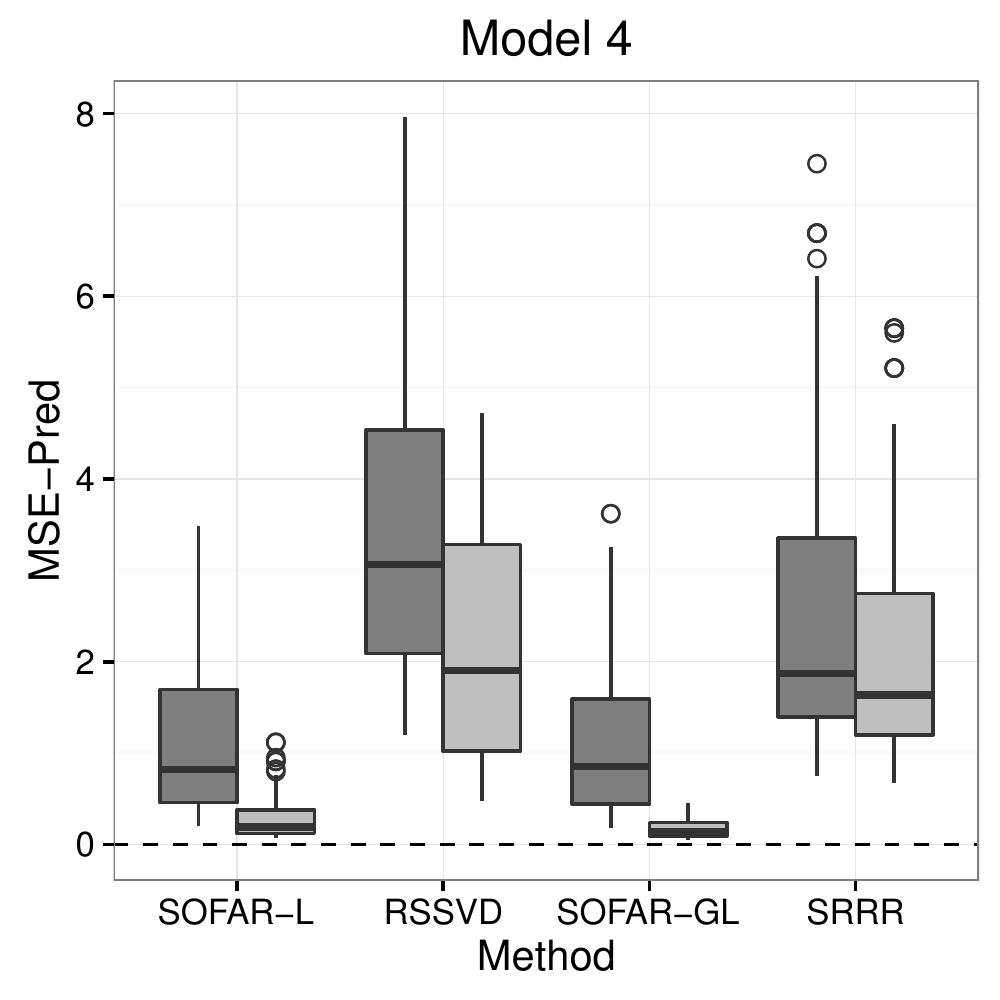}
	\caption{\label{fig:sim1}Boxplots of \textit{MSE-Pred} for Models 2 and 4 with nonadaptive (dark gray) and adaptive (light gray) versions of various methods}
\end{figure}

We first compare the performance of nonadaptive and adaptive versions of the four sparse regularization methods. Because of the space constraint, only the results in terms of \textit{MSE-Pred} in high-dimensional models 2 and 4 are presented.
The comparisons in other model settings are similar and thus omitted. From Fig. \ref{fig:sim1}, we observe that adaptive weighting generally improves the empirical performance of each method. For this reason, we only consider the adaptive versions of these
regularization methods in other comparisons.

The comparison results with adaptive penalty for Models 1 and 2 are summarized in Table \ref{tab:sim1new}. The entrywise sparse SVD structure is exactly what the SOFAR-L and RSSVD methods aim to recover. We observe that SOFAR-L performs the best among all methods in terms of both model accuracy and sparsity recovery. Although RSSVD performs only second to SOFAR-L in Model I, it has substantially worse performance in Model 2 in terms of model accuracy. This is largely because the RSSVD method does not impose any form of orthogonality constraints, which tends to cause nonidentifiability issues and compromise its performance in high dimensions. We note further that SOFAR-GL and SRRR perform worse than SOFAR-L, since they are not intended for entrywise sparsity recovery. However, these two methods still provide remarkable improvements over the OLS and RRR methods due to their ability to eliminate irrelevant variables, and over the Lasso method due to the advantages of imposing a low-rank structure. Compared to SRRR, the SOFAR-GL method results in fewer false positives and shows a clear advantage due to response selection.

The simulation results for Models 3 and 4 are reported in Table \ref{tab:sim2new}. For the rowwise sparse SVD structure in these two models, SOFAR-GL and SRRR are more suitable than the other methods. All sparse regularization methods result in higher false negative rates than in Models 1 and 2 because of the presence of some very weak signals. In Model 3, where the matrix of right singular vectors is not sparse and the dimensionality is moderate, SOFAR-GL has a slightly worse performance compared to SRRR since response selection is unnecessary. The advantages of SOFAR are clearly seen in Model 4, where the dimension is high and many irrelevant predictors and responses coexist; SOFAR-GL performs slightly better than SOFAR-L, and both methods substantially outperform the other methods. In both models, SOFAR-L and RSSVD result in higher false negative rates, since they introduce more parsimony than necessary by encouraging entrywise sparsity in $\bU$ and $\bV$. 

We have also tried models with even higher dimensions. In Model 5, we experimented with increasing the dimensions of Model 2 to $p=1000$ and $q=400$, by adding more noise variables, i.e., appending zeros to the $\bu_j^*$ and $\bv_j^*$ vectors. Table \ref{tab:sim3new} shows that the SOFAR methods still greatly outperform the others in both estimation and sparse recovery. In contrast, RSSVD becomes unstable and inaccurate; this again shows the effectiveness of enforcing the orthogonality in high-dimensional sparse SVD recovery.




\begin{table}
\hskip-3pt\caption{\label{tab:sim1new}Simulation results for Models 1--2 with various methods\protect\footnotemark[1]}
\fbox{\def~{\phantom{0}}%
\begin{tabular}{ll*{7}{l}}
\textit{Model} & \textit{Method} & \textit{MSE-Est} & \textit{MSE-Pred} & \textit{FPR (\%)} & \textit{FNR (\%)} & \textit{Rank} & \textit{Rank (\%)} & \textit{Orth}\\
         1 &        OLS & 250.7 (129.2) & 753.8 (392.2) &        100 &          0 &            &            &            \\

           &      Lasso & 12.7 (5.9) & 80.8 (34.1) &        3.8 &          0 &            &            &            \\

           &        RRR & 14.7 (6.8) & 58.6 (29.3) &        100 &          0 &          3 &        100 &          0 \\

           &    SOFAR-L &  0.4 (0.1) &  2.8 (1.3) &          0 &          0 &          3 &        100 &          0 \\

           &      RSSVD &  0.5 (0.3) &  3.8 (2.3) &        0.2 &          0 &          3 &       99.7 &        1.9 \\

           &   SOFAR-GL &  1.2 (0.5) &  8.2 (4.1) &        9.8 &          0 &          3 &        100 &          0 \\

           &       SRRR &  3.2 (1.0) & 25.2 (12.6) &       35.5 &          0 &          3 &        100 &        5.1 \\
\hline
        2 &        OLS & 1013.0 (117.0) & 765.6 (407.2) &        100 &          0 &            &            &            \\

           &      Lasso & 21.3 (7.0) & 59.0 (18.1) &        1.3 &          0 &            &            &            \\

           &        RRR & 756.4 (56.8) & 30.2 (15.9) &        100 &          0 &          3 &          0 &          0 \\

           &    SOFAR-L &  0.2 (0.1) &  0.7 (0.3) &          0 &          0 &          3 &          0 &          0 \\

           &      RSSVD &  2.5 (2.4) &  5.3 (4.1) &          1 &        0.1 &          3 &          0 &       28.4 \\

           &   SOFAR-GL &  0.7 (0.4) &  2.0 (1.0) &        2.7 &          0 &          3 &          0 &          0 \\

           &       SRRR &  3.8 (1.5) &   12.0 (6.3) &       19.8 &          0 &          3 &          0 &       40.2 \\
\end{tabular}}\\
\makeatletter
\parbox{\@tempdima}{\vskip4pt\footnotemark[1]Adaptive versions of Lasso, SOFAR-L, RSSVD, SOFAR-GL, and SRRR were applied. Means of performance measures with standard deviations in parentheses over 300 replicates are reported. \textit{MSE-Est} values are scaled by multiplying $10^4$ in Model 1 and $10^5$ in Model 2, and \textit{MSE-Pred} values are scaled by multiplying $10^3$.}
\makeatother
\end{table}

\begin{table}
\hskip-3pt\caption{\label{tab:sim2new}Simulation results for Models 3--4 with various methods\protect\footnotemark[1]}
\fbox{\def~{\phantom{0}}%
\begin{tabular}{ll*{7}{l}}
\textit{Model} & \textit{Method} & \textit{MSE-Est} & \textit{MSE-Pred} & \textit{FPR (\%)} & \textit{FNR (\%)} & \textit{Rank} & \textit{Rank (\%)} & \textit{Orth}\\
       3 &        OLS & 599.2 (339.2) & 1530.1 (870.8) &        100 &          0 &            &            &            \\

           &      Lasso & 97.6 (50.0) & 472.8 (242.7) &       15.5 &        0.6 &            &            &            \\

           &        RRR & 102.6 (70.2) & 291.9 (191.8) &        100 &          0 &          3 &        100 &          0 \\

           &    SOFAR-L & 24.8 (15.3) & 129.5 (83.2) &        0.3 &        7.4 &        3.7 &       30.3 &          0 \\

           &      RSSVD & 17.3 (11.3) & 96.6 (66.4) &        0.6 &         11 &          3 &        100 &         29 \\

           &   SOFAR-GL & 16.6 (11.4) & 94.4 (67.5) &        0.4 &        1.1 &        3.6 &       41.7 &          0 \\

           &       SRRR & 11.0 (6.7) & 63.1 (40.2) &        0.6 &        0.3 &          3 &        100 &       14.8 \\
\hline
        4 &        OLS & 252.3 (78) & 126.5 (65.4) &        100 &          0 &            &            &            \\

           &      Lasso & 37.4 (11.8) & 73.2 (24.1) &        0.8 &        2.5 &            &            &            \\

           &        RRR & 186.6 (51.6) &  6.1 (3.9) &        100 &          0 &          3 &         99 &          0 \\

           &    SOFAR-L &  0.1 (0.1) &  0.3 (0.2) &        0.1 &        4.8 &          3 &       92.7 &        0.1 \\

           &      RSSVD &    1.0 (0.7) &  2.2 (1.3) &        0.3 &       11.5 &          3 &        100 &       40.1 \\

           &   SOFAR-GL &    0.1 (0.0) &  0.2 (0.1) &          0 &        0.1 &          3 &        100 &          0 \\

           &       SRRR &  0.8 (0.5) &    2.0 (1.2) &       24.9 &        0.2 &          3 &        100 &       31.3 \\
\end{tabular}}\\
\makeatletter
\parbox{\@tempdima}{\vskip4pt\footnotemark[1]Adaptive versions of Lasso, SOFAR-L, RSSVD, SOFAR-GL, and SRRR were applied. Means of performance measures with standard deviations in parentheses over 300 replicates are reported. \textit{MSE-Est} values are scaled by multiplying $10^4$ in Model 3 and $10^5$ in Model 4, and \textit{MSE-Pred} values are scaled by multiplying $10^3$.}
\makeatother
\end{table}

\begin{table}
\hskip-3pt\caption{\label{tab:sim3new}Simulation results for Model 5. We use Model 2 with increased dimensions $p=1000$, $q=400$ by adding noise variables\protect\footnotemark[1]}
\fbox{\def~{\phantom{0}}%
\begin{tabular}{ll*{7}{l}}
\textit{Model} & \textit{Method} & \textit{MSE-Est} & \textit{MSE-Pred} & \textit{FPR (\%)} & \textit{FNR (\%)} & \textit{Rank} & \textit{Rank (\%)} & \textit{Orth}\\
         5 &        OLS & 151.5 (5.7) & 230.1 (122.9) &        100 &          0 &            &            &            \\

           &      Lasso &  3.9 (1.8) & 29.3 (11.8) &        0.6 &          0 &            &            &            \\

           &        RRR & 146.8 (7.7) & 61.5 (77.1) &        100 &          0 &        2.6 &       57.7 &          0 \\

           &    SOFAR-L &  0.1 (0.0) &  0.1 (0.0) &          0 &          0 &          3 &        100 &          0 \\

           &      RSSVD & 6.6 (14.4) &  2.8 (2.7) &        3.1 &          1 &          3 &         99 &       49.1 \\

           &   SOFAR-GL &  0.1 (0.0) &  0.2 (0.1) &        0.8 &          0 &          3 &        100 &          0 \\

           &       SRRR &  0.5 (0.2) &  3.6 (1.8) &       19.7 &          0 &          3 &        100 &       55.5 \\
\end{tabular}}\\
\makeatletter
\parbox{\@tempdima}{\vskip4pt\footnotemark[1]Adaptive versions of Lasso, SOFAR-L, RSSVD, SOFAR-GL, and SRRR were applied. Means of performance measures with standard deviations in parentheses over 300 replicates are reported. \textit{MSE-Est} values are scaled by $10^5$ and \textit{MSE-Pred} values are scaled by multiplying $10^3$.}
\makeatother
\end{table}



\subsection{Real data analysis} \label{Sec6}
In genetical genomics experiments, gene expression levels are treated as quantitative traits in order to identify expression quantitative trait loci (eQTLs) that contribute to phenotypic variation in gene expression. The task can be regarded as a multivariate regression problem with the gene expression levels as responses and the genetic variants as predictors, where both responses and predictors are often of high dimensionality. Most existing methods for eQTL data analysis exploit entrywise or rowwise sparsity of the coefficient matrix to identify individual genetic effects or master regulators \citep{Peng:Zhu:Berg:Han:Noh:regu:2010}, which not only tends to suffer from low detection power for multiple eQTLs that combine to affect a subset of gene expression traits, but also may offer little information about the functional grouping structure of the genetic variants and gene expressions. By exploiting a sparse SVD structure, the SOFAR method is particularly appealing for such applications, and may provide new insights into the complex genetics of gene expression variation.

We illustrate our approach by the analysis of a yeast eQTL data set described by \citet{Brem:Krug:land:2005}, where $n=112$ segregants were grown from a cross between two budding yeast strains, BY4716 and RM11-1a. For each of the segregants, gene expression was profiled on microarrays containing 6216 genes, and genotyping was performed at 2957 markers. Similar to \citet{Yin:Li:spar:2011}, we combined the markers into blocks such that markers with the same block differed by at most one sample, and one representative marker was chosen from each block; a marginal gene--marker association analysis was then performed to identify markers that are associated with the expression levels of at least two genes with a $p$-value less than 0.05, resulting in a total of $p=605$ markers.

Owing to the small sample size and weak genetic perturbations, we focused our analysis on $q=54$ genes in the yeast MAPK signaling pathways \citep{Kane:Goto:Sato:Kawa:Furu:data:2014}. We then applied the RRR, SOFAR-L, and SOFAR-GL methods, where adaptive weighting was used in SOFAR. We omitted the RSSVD and SRRR methods, since they do not produce a sparse SVD that obeys the orthogonality constraints.

\begin{figure}
\includegraphics[width=.24\textwidth]{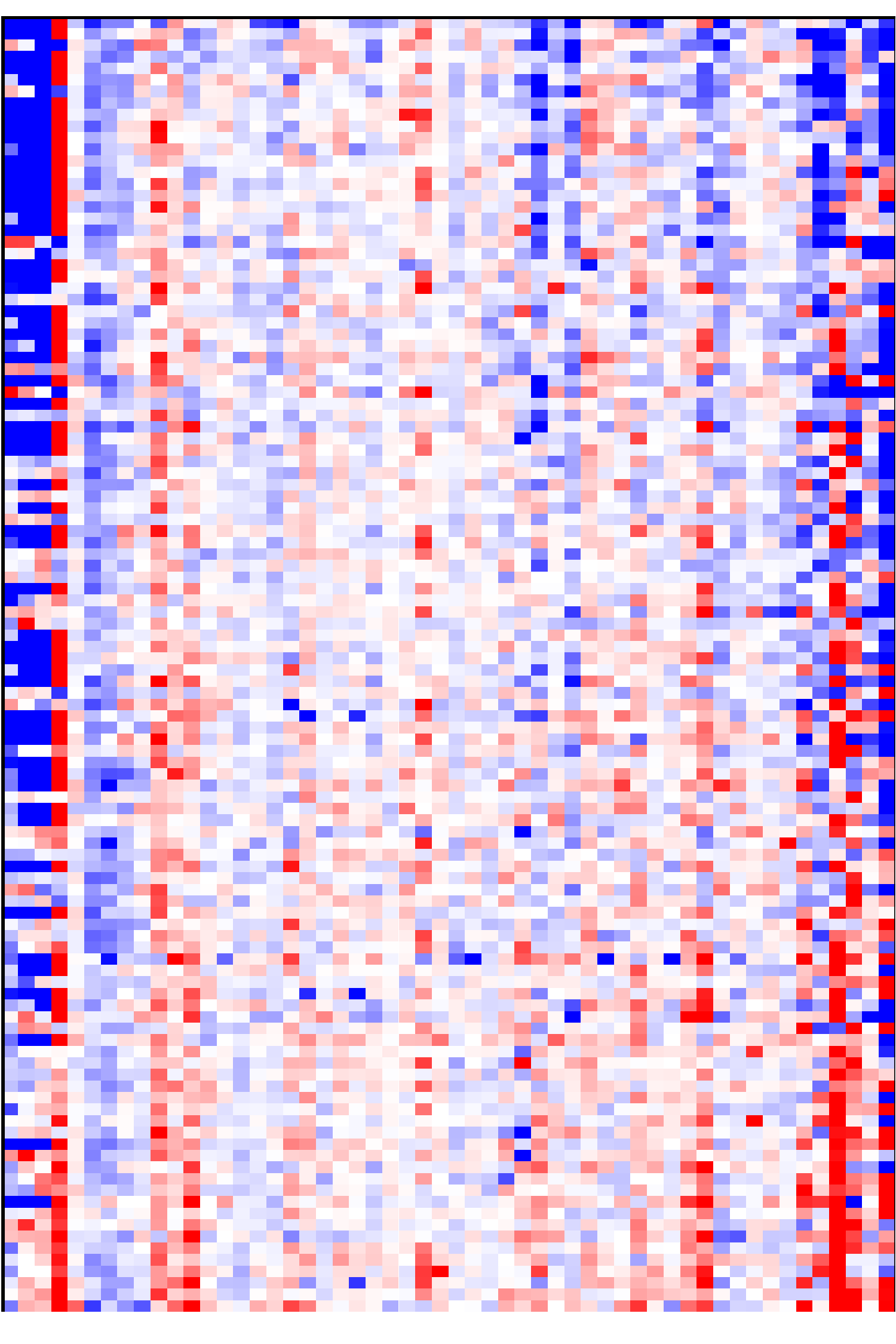}\hfill\includegraphics[width=.24\textwidth]{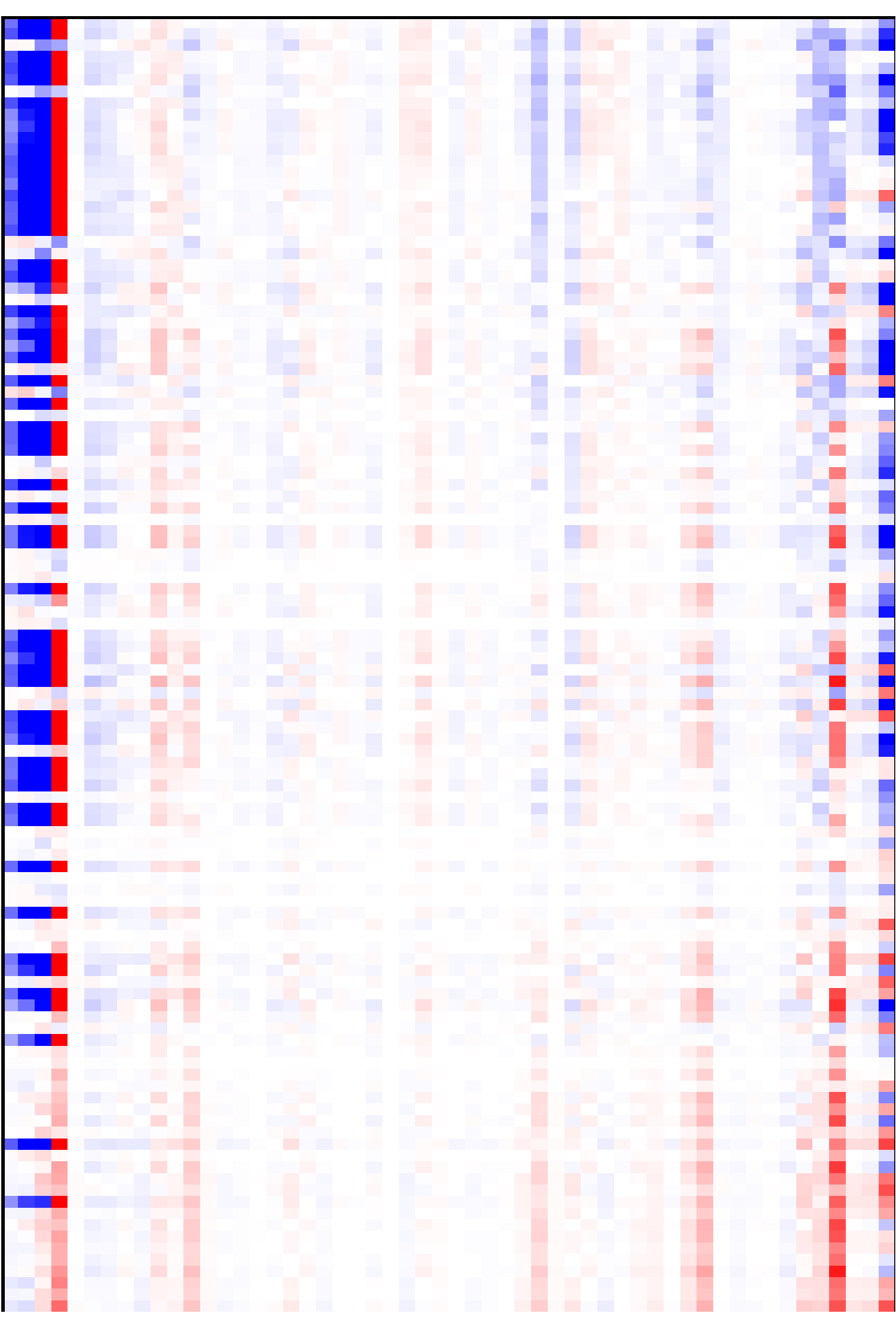}\hfill
\includegraphics[width=.24\textwidth]{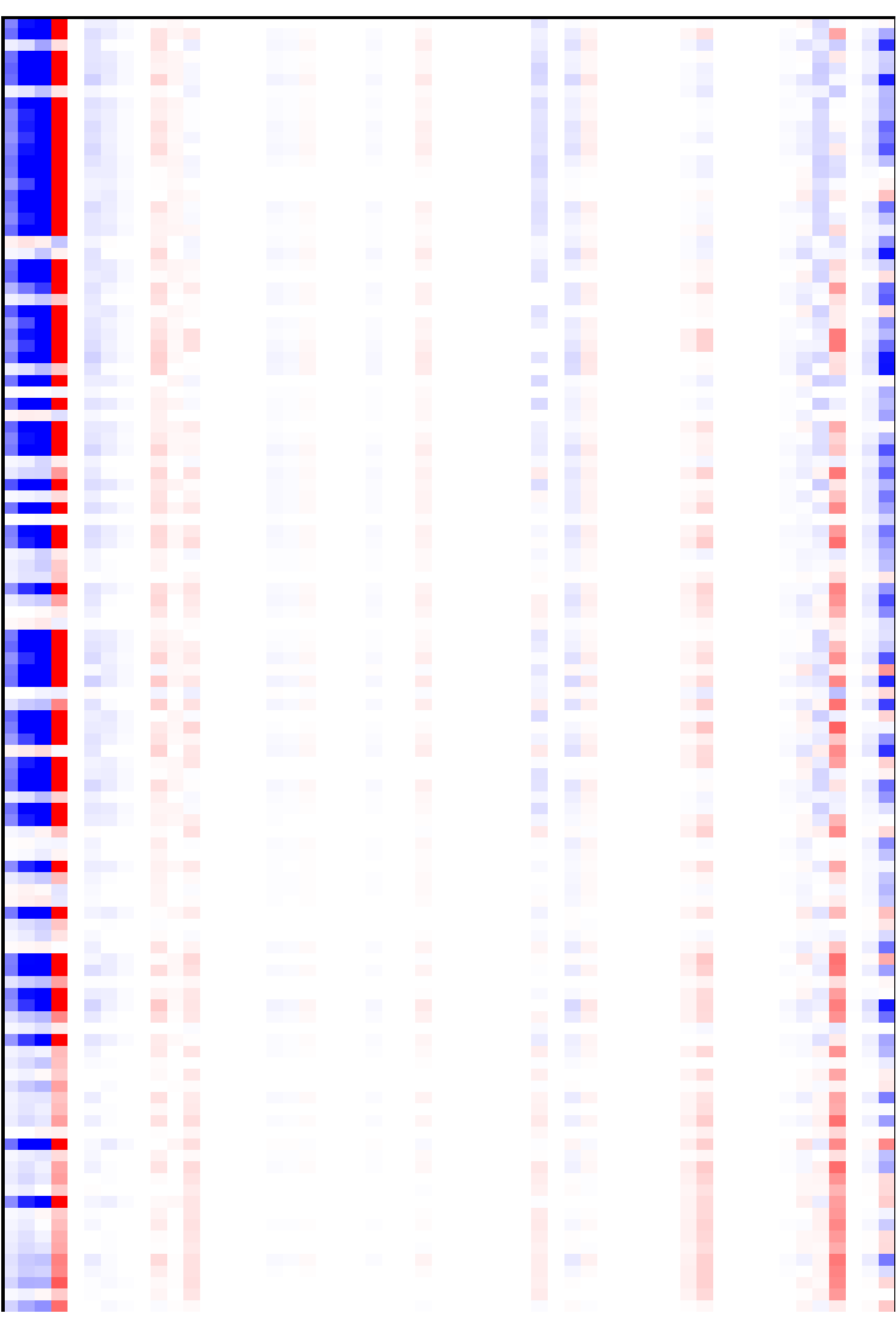}\hfill\includegraphics[width=.24\textwidth]{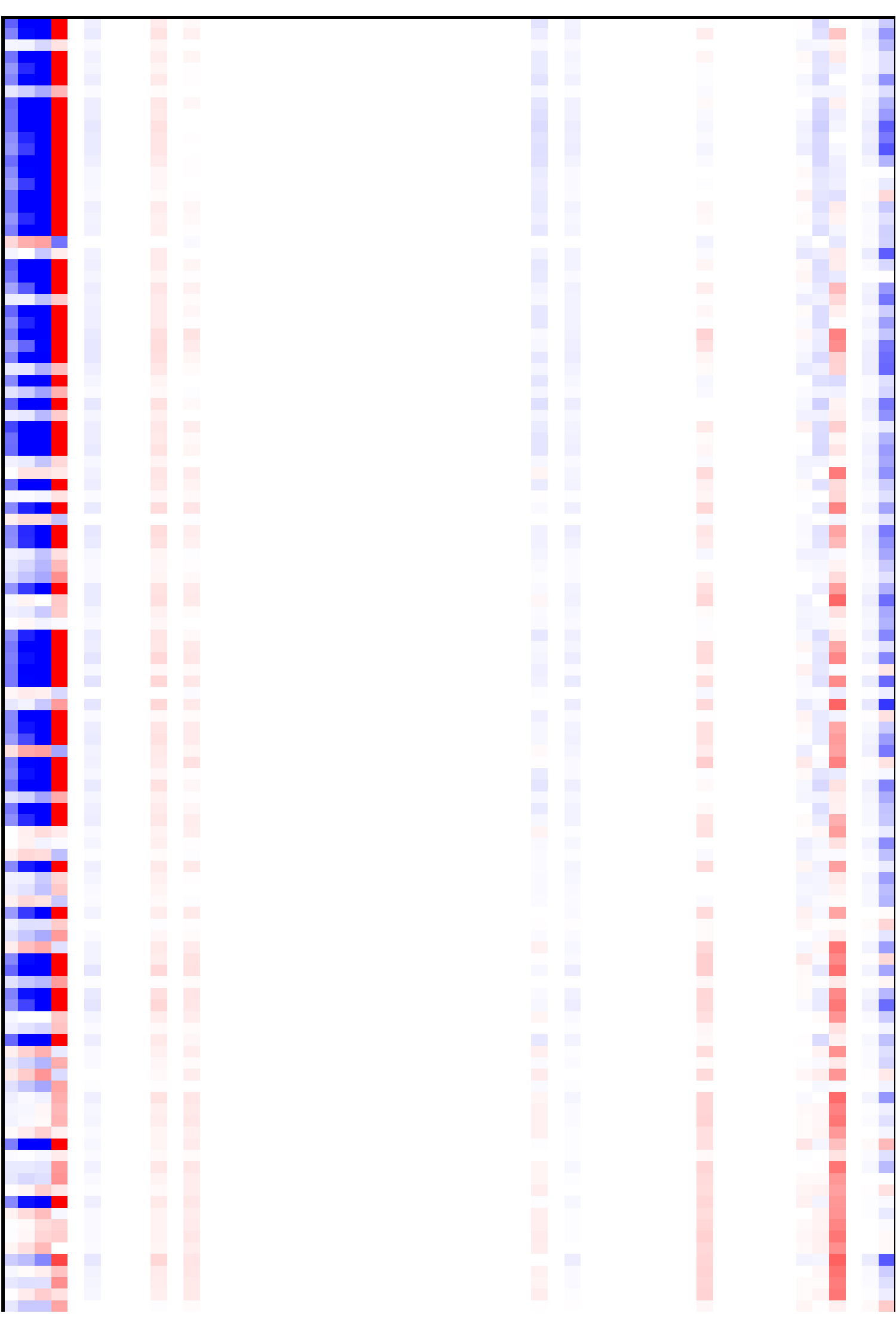}
\caption{\label{fig:heat}Heat maps of $\bY$ and its estimates by RRR, SOFAR-L, and SOFAR-GL (from left to right)}
\end{figure}

Each of the RRR, SOFAR-L, and SOFAR-GL methods resulted in a model of rank 3, indicating that dimension reduction is very effective for the data set. Also, the SVD layers estimated by the SOFAR methods are indeed sparse. The SOFAR-L estimates include 140 nonzeros in $\what\bU$, which involve only 112 markers, and 40 nonzeros in $\what\bV$, which involve only 27 genes. The sparse SVD produced by SOFAR-GL involves only 34 markers and 15 genes. The SOFAR-GL method is more conservative since it tends to identify markers that regulate all selected genes rather than a subset of genes involved in a specific SVD layer. We compare the original gene expression matrix $\bY$ and its estimates $\bX\what\bC$ by various methods using heat maps in Fig.~\ref{fig:heat}.
It is seen that the SOFAR methods achieve both low-rankness and sparsity, while still capturing main patterns in the original matrix.

\begin{figure}
\includegraphics[width=\textwidth]{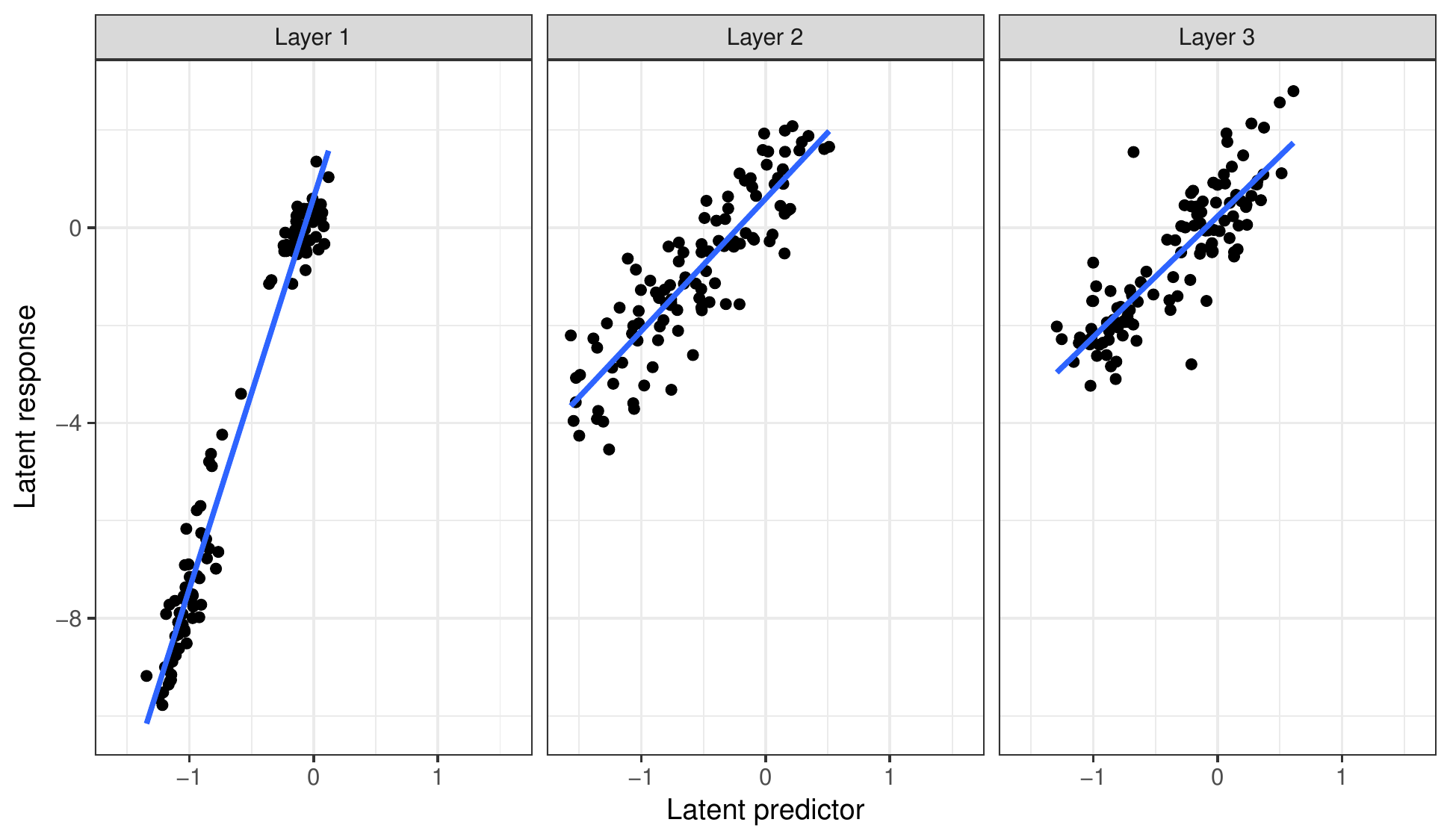}
\caption{\label{fig:layer}Scatterplots of the latent responses versus the latent predictors in three SVD layers for the yeast data estimated by the SOFAR-L method}
\end{figure}

Fig.~\ref{fig:layer} shows the scatterplots of the latent responses $\bY\what\bv_j$ versus the latent predictors $\bX\what\bu_j$ for $j=1,2,3$, where $\what\bu_j$ and $\what\bv_j$ are the $j$th columns of $\what\bU$ and $\what\bV$, respectively.
The plots demonstrate a strong association between each pair of latent variables, with the association strength descending from layer 1 to layer 3. A closer look at the SVD layers reveals further information about clustered samples and genes. The plot for layer 1 indicates that the yeast samples form two clusters, suggesting that our method may be useful for classification based on the latent variables. Also, examining the nonzero entries in $\what\bv_1$ shows that this layer is dominated by four genes, namely, STE3 ($-0.66$), STE2 ($0.59$), MFA2 ($0.40$), and MFA1 ($0.22$). All four genes are upstream in the pheromone response pathway, where MFA2 and MFA1 are genes encoding mating pheromones and STE3 and STE2 are genes encoding pheromone receptors \citep{Chen:Thor:func:2007}. The second layer is mainly dominated by CTT1 ($-0.93$), and other leading genes include SLN1 ($0.16$), SLT2 ($-0.14$),  MSN4 ($-0.14$), and GLO1 ($-0.13$). Interestingly, CTT1, MSN4, and GLO1 are all downstream genes linked to the upstream gene SLN1 in the high osmolarity/glycerol pathway required for survival in response to hyperosmotic stress. Finally, layer 3 includes the leading genes FUS1 ($0.81$), FAR1 (0.32), STE2 (0.25), STE3 (0.24), GPA1 (0.22), FUS3 (0.18), and STE12 (0.11). These genes consist of two major groups that are downstream (FUS1, FAR1, FUS3, and STE12) and upstream (STE2, STE3, and GPA1) in the pheromone response pathway. Overall, our results suggest that there are common genetic components shared by the expression traits of the clustered genes and clear reveal strong associations between the upstream and downstream genes on several signaling pathways, which are consistent with the current functional understanding of the MAPK signaling pathways.

To examine the predictive performance of each method, we randomly split the data into a training set of size 92 and a test set of size 20. The model was fitted using the training set and the predictive accuracy was evaluated on the test set based on the prediction error $\|\bY-\bX\what\bC\|_F^2/(nq)$. The splitting process was repeated 50 times. The scaled prediction errors for the RRR, SOFAR-L, SOFAR-GL, and SRRR methods are 3.3 (0.2), 2.6 (0.2), 2.4 (0.2), and 2.7 (0.1), respectively. The comparison shows the advantages of sparse and low-rank estimation. Although the SRRR method yielded similar predictive accuracy comparing to SOFAR methods on this data set, it resulted in a less parsimonious model and cannot be used for gene selection or clustering.

\bibliographystyle{rss}
\bibliography{sofar}


\def\bA{\mathbf{A}}
\def\bB{\mathbf{B}}
\def\bC{\mathbf{C}}
\def\bD{\mathbf{D}}
\def\bE{\mathbf{E}}
\def\bF{\mathbf{F}}
\def\bG{\mathbf{G}}
\def\bH{\mathbf{H}}
\def\bI{\mathbf{I}}
\def\bK{\mathbf{K}}
\def\bL{\mathbf{L}}
\def\bM{\mathbf{M}}
\def\bU{\mathbf{U}}
\def\bV{\mathbf{V}}
\def\bW{\mathbf{W}}
\def\bX{\mathbf{X}}
\def\bY{\mathbf{Y}}
\def\bZ{\mathbf{Z}}
\def\ba{\mathbf{a}}
\def\bb{\mathbf{b}}
\def\be{\mathbf{e}}
\def\bff{\mathbf{f}}
\def\bg{\mathbf{g}}
\def\bm{\mathbf{m}}
\def\bu{\mathbf{u}}
\def\bv{\mathbf{v}}
\def\bw{\mathbf{w}}
\def\bx{\mathbf{x}}
\def\by{\mathbf{y}}
\def\cA{\mathcal{A}}
\def\cB{\mathcal{B}}
\def\cC{\mathcal{C}}
\def\cK{\mathcal{K}}
\def\cS{\mathcal{S}}
\def\bdelta{\boldsymbol{\delta}}
\def\balpha{\boldsymbol{\alpha}}
\def\bGamma{\boldsymbol{\Gamma}}
\def\bDelta{\boldsymbol{\Delta}}
\def\bLambda{\boldsymbol{\Lambda}}
\def\bTheta{\boldsymbol{\Theta}}
\def\bOmega{\boldsymbol{\Omega}}
\def\bSigma{\boldsymbol{\Sigma}}
\def\blambda{\boldsymbol{\lambda}}
\def\ve{\varepsilon}
\def\bve{\boldsymbol{\varepsilon}}
\def\bzero{\mathbf{0}}
\def\bone{\mathbf{1}}
\def\what{\widehat}
\def\wtilde{\widetilde}
\newcommand{\vvec}{\mathop{\rm vec}\nolimits}
\newcommand{\supp}{\mathop{\rm supp}\nolimits}
\newcommand{\E}{\mathop{\rm E}\nolimits}

\newpage

\appendix
\setcounter{page}{1}
\setcounter{section}{0}
\renewcommand{\theequation}{A.\arabic{equation}}
\setcounter{equation}{0}

\quad \vspace{0.05in}

\begin{center}
\textbf{\Large Supplementary material to ``SOFAR: large-scale association network learning"
}

\bigskip
\author{Yoshimasa Uematsu$^1$, Yingying Fan$^1$, Kun Chen$^2$, Jinchi Lv$^1$ and Wei Lin$^3$}

\smallskip
\textit{University of Southern California$^1$, University of Connecticut$^2$ and Peking University$^3$}
\end{center}

\medskip

\noindent This Supplementary Material contains the proofs of Theorems \ref{thm:initial}--\ref{thm:conv} and additional technical details.

\section{Proofs of main results} \label{secA}

To ease the technical presentation, we introduce some necessary notation. Recall that $\bA^*=\bU^*\bD^*$, $\bB^*=\bV^*\bD^*$, $\bA=\bU\bD$, and $\bB=\bV\bD$. Denote by $\what{\bDelta} = \what{\bC} - \bC^*$, $\what{\bDelta}^d=\what{\bD}-\bD^*$, $\what{\bDelta}^a=\what{\bA}-\bA^*$, and $\what{\bDelta}^b=\what{\bB}-\bB^*$ the different estimation errors, and
$
\text{FS}(\what{\bM}) = |\{(i,j): \sgn (\what{m}_{ij}) \not= \sgn(m_{ij}^*)\}|
$ the total number of falsely discovered signs of an estimator $\what{\bM} = (\what{m}_{ij})$ for matrix $\bM^* = (m_{ij}^*)$. For $\bD=\diag(d_1,\dots,d_m)\in\mathbb{R}^{m\times m}$, we define $\bD^-$ as a  diagonal matrix with
$\rank(\bD^-)=\rank(\bD)$ and $j$th diagonal entry $d_j^- = d_j^{-1}1\{d_j>0\}$,
and define $\bD^{*-}$ based on $\bD^*$ similarly. For any matrices $\bM_1$ and $\bM_2$, denote by $\langle\bM_1, \bM_2\rangle = \tr(\bM_1^T\bM_2)$. Hereafter we use $c$ to denote a generic positive constant whose value may vary from line to line.

\subsection{Proof of Theorem \ref{thm:initial}} \label{secB.2}

We prove the bounds in \eqref{eq: Lasso-err}--\eqref{eq: ini-AB-err} separately. Recall that $s=\|\bC^*\|_0$ and define a space
\begin{align*}
\mathcal{C}_0 &= \{ \bM \in \mathbb{R}^{p\times q}:
m_{ij}=0 ~\mbox{for}~ (i,j)\not=S \},
\end{align*}
where $S$ stands for the support of $\bC^*$.
We also denote by $\mathcal{C}_0^\perp$ the orthogonal complement
of $\mathcal{C}_0$.

\medskip

\noindent \textbf{Part 1: Proof of bound \eqref{eq: Lasso-err}.} The proof is composed of two steps. We first derive the {\it deterministic} error bound \eqref{eq: Lasso-err}
under the assumption that
\begin{align}
\|n^{-1}\bX^T\bE\|_\infty \leq \lambda_0/2  \label{lem1:lamb}
\end{align}
holds almost surely in the first step and then verify that condition  (\ref{lem1:lamb}) holds
with high probability in the second step. 

\medskip

\noindent \textit{Step 1}. Since the objective function is convex, the global optimality of $\wtilde{\bC}$ implies
\begin{align*}
(2n)^{-1} \| \bY - \bX \widetilde{\bC} \|_F^2+ \lambda_0 \| \widetilde{\bC} \|_1
\leq
(2n)^{-1} \| \bY - \bX \bC^* \|_F^2 + \lambda_0 \| \bC^* \|_1.
\end{align*}
Then letting $\widetilde{\bDelta}\equiv \wtilde{\bC}-\bC^*$, we see that
\begin{align}
(2n)^{-1} \|\bX \widetilde{\bDelta} \|_F^2
\leq \langle n^{-1} \bX^T\bE, \widetilde{\bDelta} \rangle
+\lambda_0 ( \| {\bC}^* \|_1 - \| \widetilde{\bDelta} + \bC^* \|_1). \label{iniest01}
\end{align}
By H\"{o}lder's inequality and the assumed condition (\ref{lem1:lamb}), it holds that
\begin{align}
\langle n^{-1} \bX^T\bE, \widetilde{\bDelta} \rangle
\leq \| n^{-1} \bX^T\bE \|_\infty \| \widetilde{\bDelta} \|_1
\leq 2^{-1}\lambda_0 \| \widetilde{\bDelta} \|_1. \label{iniest02}
\end{align}
By the triangle inequality, we have
\begin{align}
\lambda_0 ( \| {\bC}^* \|_1 - \| \widetilde{\bDelta} + \bC^* \|_1)
\leq \lambda_0 \| \widetilde{\bDelta} \|_1.  \label{iniest04}
\end{align}
Therefore, (\ref{iniest01}) together with Lemma \ref{lem:rsc} in Section \ref{secB.1}  and
(\ref{iniest02})--(\ref{iniest04}) entails that
\begin{align}
2c_2 \|\widetilde{\bDelta} \|_F^2 \leq 2n^{-1}\|\bX \widetilde{\bDelta} \|_F^2
\leq
6\lambda_0 \| \widetilde{\bDelta} \|_1. \label{iniest05}
\end{align}

Meanwhile, since $n^{-1}\|\bX \widetilde{\bDelta} \|_F^2$ is nonnegative
(\ref{iniest01}) is also bounded from below as
\begin{align}
0\leq
\langle n^{-1} \bX^T\bE, \widetilde{\bDelta} \rangle
+ \lambda_0 ( \| {\bC}^* \|_1 - \| \widetilde{\bDelta} + \bC^* \|_1)
. \label{iniest06}
\end{align}
Note that $\bC_{\mathcal{C}_0^\perp}^*=\bzero$ in our model.
Hence it follows from the triangle inequality and decomposability of the nuclear norm that
\begin{align}
\lambda_0 ( \| {\bC}^* \|_1  - \|\widetilde{\bDelta} + \bC^* \|_1 )
&= \lambda_0 (\| \bC_{\mathcal{C}_0}^* + \bC_{\mathcal{C}_0^\perp}^* \|_1
- \| \widetilde{\bDelta}_{\mathcal{C}_0} + \widetilde{\bDelta}_{\mathcal{C}_0^\perp}
+ \bC_{\mathcal{C}_0}^* + \bC_{\mathcal{C}_0^\perp}^* \|_1 ) \notag \\
&\leq \lambda_0 (\| \bC_{\mathcal{C}_0}^* \|_1 + \| \bC_{\mathcal{C}_0^\perp}^* \|_1
- \| \bC_{\mathcal{C}_0}^*
+ \widetilde{\bDelta}_{\mathcal{C}_0^\perp} \|_1
+ \| \bC_{\mathcal{C}_0^\perp}^*
+ \widetilde{\bDelta}_{\mathcal{C}_0} \|_1 ) \notag \\
&= \lambda_0 ( \| \widetilde{\bDelta}_{\mathcal{C}_0} \|_1
- \| \widetilde{\bDelta}_{\mathcal{C}_0^\perp} \|_1). \label{iniest07}
\end{align}
Thus by (\ref{iniest02}) and (\ref{iniest07}), we can bound (\ref{iniest06}) from above as
\begin{align}
0 &\leq
2^{-1}\lambda_0 \| \widetilde{\bDelta} \|_1
+ \lambda_0 ( \| \widetilde{\bDelta}_{\mathcal{C}_0} \|_1
- \| \widetilde{\bDelta}_{\mathcal{C}_0^\perp} \|_1) \notag \\
&\leq 2^{-1}\lambda_0 ( \| \widetilde{\bDelta}_{\mathcal{C}_0} \|_1 + \| \widetilde{\bDelta}_{\mathcal{C}_0^\perp} \|_1)
+ \lambda_0 ( \| \widetilde{\bDelta}_{\mathcal{C}_0} \|_1- \| \widetilde{\bDelta}_{\mathcal{C}_0^\perp} \|_1) \notag \\
&= 2^{-1}\lambda_0 ( 3\| \widetilde{\bDelta}_{\mathcal{C}_0} \|_1-\| \widetilde{\bDelta}_{\mathcal{C}_0^\perp} \|_1), \notag
\end{align}
which can be equivalently rewritten as
\begin{align}
\lambda_0\| \widetilde{\bDelta}_{\mathcal{C}_0^\perp} \|_1
\leq
3\lambda_0  \| \widetilde{\bDelta}_{\mathcal{C}_0} \|_1. \label{iniest09}
\end{align}

We are now ready to derive the error bound.
For a generic positive constant $c$, (\ref{iniest05}) is bounded from above by the decomposability of the $\ell_1$-norm
and (\ref{iniest09}) as
\begin{align}
c \|\widetilde{\bDelta} \|_F^2
\leq \lambda_0 \| \widetilde{\bDelta} \|_1
= \lambda_0 \| \widetilde{\bDelta}_{\mathcal{C}_0} \|_1+ \lambda_0 \| \widetilde{\bDelta}_{\mathcal{C}_0^\perp} \|_1
\leq 4 \lambda_0 \| \widetilde{\bDelta}_{\mathcal{C}_0} \|_1. \label{iniest10}
\end{align}
Using the subspace compatibility conditions (see the proof of Theorem 1 of
\cite{Negahbanetal2012}), we can show that
\begin{align*}
\| \widetilde{\bDelta}_{\mathcal{C}_0} \|_1
\leq s^{1/2}\| \widetilde{\bDelta}_{\mathcal{C}_0} \|_F
\leq s^{1/2}\|\widetilde{\bDelta}\|_F.
\end{align*}
Therefore, with $c$ changed appropriately (\ref{iniest10}) can be further bounded as
\begin{align*}
\|\widetilde{\bDelta} \|_F^2
\leq  cs^{1/2}\lambda_0 \|\widetilde{\bDelta}\|_F.
\end{align*}
This consequently yields the desired error bound
\begin{align*}
\|\widetilde{\bDelta} \|_F
\leq c s^{1/2} \lambda_0,
\end{align*}
which completes the first step of the proof.

\medskip

\noindent \textit{Step 2}. Let $\bx_i$ and $\be_j$ denote
the $i$th and $j$th columns of $\bX \in \mathbb{R}^{n\times p}$ and
$\bE \in \mathbb{R}^{n\times q}$, respectively.
Since $\|\bX^T \bE \|_\infty=\max_{1\leq i \leq p}\max_{1\leq j \leq q}|\bx_i^T \be_j|$, using Bonferroni's inequality and
the Gaussianity of $\be_j$ we deduce
\begin{align} \label{neweq012}
P\left( n^{-1}\|\bX^T \bE \|_\infty \geq \lambda_0 \right)
& \leq \sum_{i=1}^p\sum_{j=1}^q P\left( n^{-1}|\bx_i^{T}\be_j| \geq \lambda_0 \right) \notag\\
&
\leq 2\sum_{i=1}^p\sum_{j=1}^q \exp \left( -\frac{n^2\lambda_0^2}{2\mathbb{E} |\bx_i^{T} \be_j|^2} \right).
\end{align}
Since $\be_j$ is distributed as $N\left(0,\sigma_j^2 \bI_n\right)$, it holds that
\begin{equation} \label{neweq013}
\mathbb{E} |\bx_i^{T}\be_j|^2 = \sigma_j^2 \bx_i^{T}\bx_i
\leq \sigma_{\max}^2 n.
\end{equation}
By the assumption $\lambda_0^2 = c_0^2\sigma_{\max}^2 n^{-1}\log (pq) $ and (\ref{neweq012})--(\ref{neweq013}),
the upper bound on the probability in \eqref{neweq012} can be further bounded from above by
\begin{align*}
2pq \exp \left\{ -(c_0^2/2)\log (pq) \right\}
= 2(pq)^{1-c_0^2/2},
\end{align*}
which concludes the proof for bound \eqref{eq: Lasso-err}.

\medskip

\noindent \textbf{Part 2: Proofs of bounds \eqref{eq: ini-D-err} and \eqref{eq: ini-AB-err}}. Both inequalities \eqref{eq: ini-D-err} and \eqref{eq: ini-AB-err} are direct consequences of Lemma \ref{lem:MPT} in Section \ref{secB.3} and bound \eqref{eq: Lasso-err}. This completes the proof of Theorem \ref{thm:initial}.

\subsection{Proof of Theorem \ref{thm:det}} \label{secA.1}

Recall that we solve SOFAR in a local neighborhood $\mathcal{P}_n$ of the initial solution $\wtilde{\bC}$. It follows that $\|\what\bDelta\|_F \leq \|\what{\bC} - \wtilde{\bC}\|_F + \|\wtilde{\bC} - \bC^*\|_F \leq 3R_n \leq cs^{1/2}\lambda_{\max}$,  where $\mathcal{P}_n$ is defined in
(\ref{def: local-neighbor}),
$R_n$ is as in Theorem \ref{thm:initial}, and $c$ is some generic positive constant. Thus by Lemma \ref{lem:MPT}, we have
\begin{align}
 \|\what{\bDelta}^a\|_F + \|\what{\bDelta}^b\|_F + \|\what{\bDelta}^d\|_F & \leq c \eta_n\|\what{\bDelta}\|_F \label{eq:003}\\
& \leq cs^{1/2}\lambda_{\max}\eta_n, \label{eq:002}
\end{align}
where $\eta_n = 1+\delta^{-1/2} \big(\sum_{j=1}^r(d_1^*/d_j^*)^2\big)^{1/2}$.
Note that under Conditions \ref{Aparsp} and \ref{Aeigen},
Lemma \ref{lem:rsc} and Lemma \ref{lemma: l1-bound} in Section  \ref{secA.4} entail that
\begin{align}
 \|\what{\bDelta}\|_F^2 \leq c n^{-1}\|\bX\what{\bDelta}\|_F^2 \leq c \lambda_{\max}\Big(\|\what{\bDelta}^d\|_1 + \|\what{\bDelta}^a\|_1 + \|\what{\bDelta}^b\|_1\Big). \label{Thm1:05}
\end{align}
Furthermore,  it follows from the Cauchy--Schwarz inequality and \eqref{eq:003} that
\begin{align}
\nonumber &\|\what{\bDelta}^a\|_1 + \|\what{\bDelta}^d\|_1 + \|\what{\bDelta}^b\|_1\\
\nonumber &\leq \max\{\|\what{\bDelta}^d\|_0, \|\what{\bDelta}^a\|_0, \|\what{\bDelta}^b\|_0\}^{1/2} \left( \|\what{\bDelta}^a\|_F + \|\what{\bDelta}^d\|_F + \|\what{\bDelta}^b\|_F \right)\\
& \leq c \eta_n\{ \|\what{\bDelta}^d\|_0 + \|\what{\bDelta}^a\|_0 + \|\what{\bDelta}^b\|_0\}^{1/2}  \|\what{\bDelta}\|_F. \label{eq: l1-l0}
\end{align}
Combining \eqref{eq: l1-l0} and \eqref{Thm1:05} leads to
\begin{align}
\|\what{\bDelta}\|_F
& \leq c \lambda_{\max}\eta_n\{ \|\what{\bDelta}^d\|_0 + \|\what{\bDelta}^a\|_0 + \|\what{\bDelta}^b\|_0\}^{1/2}.\label{eq:005}
\end{align}

We next provide an upper bound for $\|\what{\bDelta}^d\|_0 + \|\what{\bDelta}^a\|_0 + \|\what{\bDelta}^b\|_0$.  Since  $(\what{\bD},\what{\bA},\what{\bB})$ and
$({\bD}^*,{\bA}^*,{\bB}^*)$ are elements in $\mathcal{D} \times \mathcal{A} \times \mathcal{B}$
by Condition \ref{Aparsp}, we have
\begin{align}
\text{FS}(\what{\bD})^{1/2}\tau \leq \|\what{\bDelta}^d\|_F,~~~
\text{FS}(\what{\bA})^{1/2}\tau \leq \|\what{\bDelta}^a\|_F,~~~\text{ and } ~~~
\text{FS}(\what{\bB})^{1/2}\tau \leq \|\what{\bDelta}^b\|_F.  \label{Thm1:10}
\end{align}
By the definition of $\text{FS}(\what{\bA})$, it holds that
$\|\what{\bDelta}^a\|_0 \leq s_a + \text{FS}(\what{\bA})$.
Similar inequalities hold for $\|\what{\bDelta}^b\|_0$ and $\|\what{\bDelta}^d\|_0$.
Therefore, it follows from \eqref{Thm1:10} and \eqref{eq:003} that
\begin{align}
\nonumber \|\what{\bDelta}^d\|_0+\|\what{\bDelta}^a\|_0+\|\what{\bDelta}^b\|_0
&\leq r + s_a + s_b
+ \text{FS}(\what{\bD})+ \text{FS}(\what{\bA})+ \text{FS}(\what{\bB}) \\
\nonumber& \leq r+s_a+s_b + \tau^{-2}\Big(\|\what{\bDelta}^a\|_F + \|\what{\bDelta}^b\|_F + \|\what{\bDelta}^d\|_F\Big)^2\\
& \leq  r + s_a + s_b + c (\eta_n/\tau)^2 \|\what{\bDelta}\|_F^2. \label{eq:001}
\end{align}
Plugging \eqref{eq:001} into \eqref{eq:005} yields
\begin{align*}
\|\what{\bDelta}\|_F \leq c \lambda_{\max}\eta_n\left( r + s_a + s_b + c(\eta_n/\tau)^2 \|\what{\bDelta}\|_F^2\right)^{1/2}.
\end{align*}
Thus solving for $\|\what{\bDelta}\|_F$ gives
\begin{align}\label{eq:F-bound}
\|\what{\bDelta}\|_F \leq \frac{c(r+s_a+s_b)^{1/2}\lambda_{\max}\eta_n}{\left\{1-c\lambda_{\max}^2(\eta_n^2/\tau)^2\right\}^{1/2} },
\end{align}
which together with Theorem \ref{thm:initial} results in the first inequality in Theorem \ref{thm:det}.

Plugging \eqref{eq:F-bound} into  \eqref{eq:003}, we deduce
\begin{align*}
\|\what{\bDelta}^a\|_F + \|\what{\bDelta}^b\|_F + \|\what{\bDelta}^d\|_F & \leq  \frac{c(r+s_a+s_b)^{1/2}\lambda_{\max}\eta_n^2}{\left\{1-c\lambda_{\max}^2(\eta_n^2/\tau)^2 \right\}^{1/2} },
\end{align*}
which along with \eqref{eq:002} entails the second inequality in Theorem \ref{thm:det}. Note that plugging \eqref{eq:F-bound} into  \eqref{eq:001} and combining terms yield
\begin{align*}
\|\what{\bDelta}^d\|_0+\|\what{\bDelta}^a\|_0+\|\what{\bDelta}^b\|_0 \leq \frac{c(r+s_a+s_b)}{1-c\lambda_{\max}^2(\eta_n^2/\tau)^2},
\end{align*}
which gives the third inequality in Theorem \ref{thm:det}.

We now plug the above inequality and \eqref{eq:F-bound}  into \eqref{eq: l1-l0}. Then it holds that
\begin{align}
\|\what{\bDelta}^a\|_1 + \|\what{\bDelta}^d\|_1 + \|\what{\bDelta}^b\|_1 \leq \frac{c(r+s_a+s_b)\lambda_{\max}\eta_n^2}{1-c\lambda_{\max}^2(\eta_n^2/\tau)^2}, \label{eq: 006}
\end{align}
which yileds the fourth inequality in Theorem \ref{thm:det}. Finally, it follows from Lemma \ref{lemma: l1-bound} and \eqref{eq: 006} that
\begin{align*}
n^{-1}\|\bX\what{\bDelta}\|_F^2 \leq \frac{c(r+s_a+s_b)\lambda_{\max}^2\eta_n^2}{1-c\lambda_{\max}^2(\eta_n^2/\tau)^2},
\end{align*}
which establishes the fifth inequality in the theorem and concludes the proof of Theorem \ref{thm:det}.

\subsection{Lemma \ref{lemma: l1-bound} and its proof} \label{secA.4}

\begin{lemma} \label{lemma: l1-bound}
Under the conditions of Theorem \ref{thm:det}, with at least probability as specified in \eqref{inequalities-probability} we have
\begin{align*}
n^{-1}\|\bX\what{\bDelta}\|_F^2
\leq c\lambda_{\max} \left( \|\what{\bDelta}^d\|_1
+ \|\what{\bDelta}^a\|_1 + \|\what{\bDelta}^{b}\|_1 \right), 
\end{align*}
where $c$ is some 
positive constant.
\end{lemma}

\textit{Proof of Lemma \ref{lemma: l1-bound}}. Denote by $\mathcal{E}_2$ the event on which inequalities (\ref{A:csp1})--(\ref{A:csp3}) hold. Then by Lemma \ref{lem:inequalities} in Section \ref{secA.2}, we see that event $\mathcal{E}_2$ holds with probability bound as specified in \eqref{inequalities-probability}. We will prove Lemma \ref{lemma: l1-bound} by conditioning on event $\mathcal{E}_2$. Since the SOAR estimator is the minimizer in the neighborhood $\mathcal{P}_n$ defined in (\ref{def: local-neighbor}), 
it holds that
\begin{align*}
& (2n)^{-1}\|\bY - \bX\what{\bU}\what{\bD} \what{\bV}^T\|_F^2 + \lambda_d\|\what{\bD}\|_1 + \lambda_a\rho_a(\what{\bA}) + \lambda_b\rho_b(\what{\bB}) \\
& \leq (2n)^{-1}\|\bY - \bX\bU^*\bD^* (\bV^*)^T\|_F^2 + \lambda_d\|\bD^*\|_1 + \lambda_a\rho_a(\bA*) + \lambda_b\rho_b(\bB^*).
\end{align*}
Let $\what{\bDelta} = \what{\bC} - \bC^*$. Rearranging terms in the above inequality leads to
\begin{align}
&(2n)^{-1} \|\bX \widehat{\bDelta}\|_F^2 \leq \langle n^{-1}\bX^T \bE,\widehat{\bDelta} \rangle \notag \\
&\qquad +\lambda_d\left(\|\bD^*\|_1-\|\what{\bD}\|_1 \right)
+ \lambda_a\left(\rho_a(\bA^*)-\rho_a(\what{\bA}) \right)
+ \lambda_b\left(\rho_b(\bB^*)-\rho_b(\what{\bB}) \right). \label{Thm1:01}
\end{align}
By the definition of $\bD^-$, the estimation error can be decomposed as
\begin{align*}
\what\bDelta & \equiv \what\bU\what\bD\what\bV^T - \bU^*\bD^*\bV^{*T}
= \what \bA\what\bD^-\what\bB^T - \bA^*\bD^{*-}\bB^{*T} \\
&= \what\bDelta^a (\what\bB \what\bD^-)^T - \bU^*\what\bDelta^d (\what\bB \what\bD^-)^T + \bU^*(\what\bDelta^{b})^{T}.
\end{align*}
The above decomposition together with
H\"{o}lder's inequality entails that the following inequality
\begin{align}
&\langle n^{-1}\bX^T\bE,\widehat{\bDelta} \rangle \notag \\
&= \langle n^{-1}\bX^T \bE \what{\bB} \what{\bD}^-,\what{\bDelta}^a \rangle  
- \langle n^{-1}\bU^{*T} \bX^T \bE \what{\bB} \what{\bD}^-,\what{\bDelta}^d \rangle
+ \langle n^{-1}\bU^{*T} \bX^T \bE ,\what{\bDelta}^{bT} \rangle \notag \\
&\leq \|n^{-1}\bX^T \bE \what{\bB} \what{\bD}^-\|_\infty \|\what{\bDelta}^a\|_1
+ \|n^{-1}\bU^{*T}\bX^T\bE\what{\bB} \what{\bD}^-\|_\infty \|\what{\bDelta}^d\|_1
+ \|n^{-1}\bU^{*T}\bX^T\bE\|_\infty \|\what{\bDelta}^{b}\|_1 \notag \\
&\leq \lambda_a \|\what{\bDelta}^a\|_1
+ \lambda_d \|\what{\bDelta}^d\|_1
+ \lambda_b \|\what{\bDelta}^{b}\|_1 \label{Thm1:02}
\end{align}
holds on event $\mathcal{E}_2$.

By the triangle inequality for the $\ell_1$-norm and Condition \ref{Apen}, we deduce
\begin{align}
&\lambda_d\left(\|\bD^*\|_1-\|\what{\bD}\|_1 \right)
+ \lambda_a\left(\rho_a(\bA^*)-\rho_a(\what{\bA}) \right)
+ \lambda_b\left(\rho_b(\bB^*)-\rho_b(\what{\bB}) \right) \notag \\
&\leq \lambda_d \|\what{\bDelta}^d\|_1
+ \lambda_a \|\what{\bDelta}^a\|_1
+ \lambda_b \|\what{\bDelta}^{b}\|_1. \label{Thm1:03}
\end{align}
Thus plugging (\ref{Thm1:02}) and (\ref{Thm1:03}) into (\ref{Thm1:01}) yields
\begin{align}
(cn)^{-1}\|\bX\what{\bDelta}\|_F^2
&\leq \lambda_d \|\what{\bDelta}^d\|_1
+ \lambda_a \|\what{\bDelta}^a\|_1
+ \lambda_b \|\what{\bDelta}^{b}\|_1 \notag \\
&\leq \lambda_{\max} \left( \|\what{\bDelta}^d\|_1
+ \|\what{\bDelta}^a\|_1 + \|\what{\bDelta}^{b}\|_1 \right) \label{Thm1:04}
\end{align}
with $\lambda_{\max}=\max(\lambda_d,\lambda_a,\lambda_b)$, which completes the proof of Lemma \ref{lemma: l1-bound}.

\subsection{Lemma \ref{lem:inequalities} and its proof} \label{secA.2}

\begin{lemma}
\label{lem:inequalities}
Under the conditions of Theorem \ref{thm:det}, with at least probability as specified in \eqref{inequalities-probability} the following inequalities hold
\begin{align}
\sup_{(\bB,\bD)\in\mathcal{P}_n} \|n^{-1} \bU^{*T}\bX^T\bE \bB\bD^-\|_\infty &\leq \lambda_d,  \label{A:csp1} \\
\sup_{(\bB,\bD)\in\mathcal{P}_n} \|n^{-1} \bX^T\bE \bB\bD^-\|_\infty &\leq \lambda_a,  \label{A:csp2} \\
\sup_{(\bB,\bD)\in\mathcal{P}_n} \|n^{-1} \bU^{*T}\bX^T\bE\|_\infty &\leq \lambda_b. \label{A:csp3}
\end{align}
\end{lemma}

\textit{Proof of Lemma \ref{lem:inequalities}}. Recall that $\wtilde{\mathcal{P}}_n = \{\bC: \|\bC - \wtilde\bC\|_F \leq 2R_n\}$, where $\wtilde\bC$ is the initial Lasso estimator and  $R_n=c(n^{-1}s\log(pq))^{1/2}$ is as defined in Theorem \ref{thm:initial}. It follows from Theorem \ref{thm:initial} that the true regression coefficient matrix $\bC^*$ falls in the neighborhood $\wtilde{\mathcal{P}}_n$ with probability at least $1- 2(pq)^{1-c_0^2/2}$, where $c_0>\sqrt{2}$ is some constant given in Theorem \ref{thm:initial}. Note that the neighborhood $\wtilde{\mathcal{P}}_n$ shrinks asymptotically as $n\rightarrow \infty$ since  $R_n^2=O(n^{\alpha+\beta/2+\gamma-1})$ and $\alpha+\beta/2+\gamma<\alpha+\beta+\gamma < 1$ holds
under our assumptions. In order to deal with the nonconvexity of the objective function, we exploit the framework of convexity-assisted nonconvex optimization (CANO) and solve the SOFAR optimization problem in the shrinking local region $\mathcal{P}_n = \wtilde{\mathcal{P}}_n \cap (\mathcal{C} \times \mathcal{D} \times \mathcal{A} \times \mathcal{B})$ as defined in (\ref{def: local-neighbor}).

Observe that for any $\bC \in \wtilde{\mathcal{P}}_n$, by the triangle inequality it holds that
\[\|\bC - \bC^*\|_F \leq \|\bC-\wtilde\bC\|_F + \|\wtilde\bC - \bC^*\|_F\leq 3R_n; \]
that is, with probability at least $1- 2(pq)^{1-c_0^2/2}$,  $\wtilde{\mathcal{P}}_n \subset \{\bC: \|\bC - \bC^*\|_F \leq 3R_n\}$.  
Further, by Lemma \ref{lem:MPT} we have $\{\bC: \|\bC - \bC^*\|_F \leq 3R_n\} \subset \mathcal{E}_1$, where
\begin{align}\label{eq: ABD-bounds}
\mathcal{E}_1 = \{ & \bC \equiv \bA\bD^{-}\bB: \|\bD - \bD^*\|_F  \leq 3R_n, \nonumber \\
& \quad \|\bA - \bA^*\|_F + \|\bB - \bB^*\|_F \leq 3 c\eta_nR_n\}
\end{align}
with $c>0$ some constant. Combining the above results yields that with probability at least $1- 2(pq)^{1-c_0^2/2}$,  $\mathcal{P}_n \subset \wtilde{\mathcal{P}}_n \subset \mathcal{E}_1$, which entails
\begin{align}\label{eq:010}
P\Big(\mathcal{P}_n \not\subset \mathcal{E}_1\Big) \leq 2(pq)^{1-c_0^2/2}.
\end{align}



We next establish that (\ref{A:csp1})--(\ref{A:csp3}) hold with asymptotic probability one. Note that it follows from the definition of conditional probability and \eqref{eq:010} that
\begin{align*}
P\Big(& \sup_{\bC \in \mathcal{P}_n} \|n^{-1}\bU^*\bX^T\bE\bB\bD^{-}\|_\infty > \lambda_d \Big)\\
& \leq  P\Big(\sup_{\bC \in \mathcal{P}_n} \|n^{-1}\bU^*\bX^T\bE\bB\bD^{-}\|_\infty > \lambda_d \Big| \mathcal{P}_n \subset \mathcal{E}_1\Big) + P\Big(\mathcal{P}_n \not\subset \mathcal{E}_1\Big) \\
& \leq P\Big(\sup_{\bC \in \mathcal{E}_1} \|n^{-1}\bU^*\bX^T\bE\bB\bD^{-}\|_\infty > \lambda_d) + 2(pq)^{1-c_0^2/2}.
\end{align*}
Thus to prove \eqref{A:csp1}, we only need to show that
\begin{align}\label{A:csp1-new}
\sup_{\bC \in \mathcal{E}_1} \|n^{-1}\bU^*\bX^T\bE\bB\bD^{-}\|_\infty \leq \lambda_d
\end{align}
holds with asymptotic probability one. Similarly, to show \eqref{A:csp2} and \eqref{A:csp3} we only need to prove that
\begin{align}
\sup_{\bC\in\mathcal{E}_1} \|n^{-1} \bX^T\bE \bB\bD^-\|_\infty &\leq \lambda_a,  \label{A:csp2-new} \\
\sup_{\bC\in\mathcal{E}_1} \|n^{-1} \bU^{*T}\bX^T\bE\|_\infty &\leq \lambda_b\label{A:csp3-new}
\end{align}
hold with asymptotic probability one. We next proceed to prove \eqref{A:csp1-new}--\eqref{A:csp3-new} hold with asymptotic probability one.

Denote by $\bx_i$ and $\be_j$
the $i$th and $j$th columns of $\bX \in \mathbb{R}^{n\times p}$ and
$\bE \in \mathbb{R}^{n\times q}$, respectively.
Let $\bx_i^*$ and $\be_j^*$ be
the $i$th and $j$th columns of $\bX^*\equiv\bX\bU^* \in \mathbb{R}^{n\times q}$ and
$\bE^*\equiv\bE\bV^* \in \mathbb{R}^{n\times q}$, respectively. It is seen that the last $q-r$ columns of $\bX^*$ and $\bE^*$ are all zero.
First, we show that (\ref{A:csp1-new}) holds with significant probability.
The decomposition
\begin{align*}
\bB \bD^- = \bV^* + \bDelta^b\bD^- + \bV^*\bD^*\bDelta^{d-}
\end{align*}
and the triangle inequality lead to
\begin{align*}
\|n^{-1}\bX^{*T}\bE\bB \bD^- \|_\infty
\leq \|n^{-1}\bX^{*T}\bE^* \|_\infty
+ \|n^{-1}\bX^{*T}\bE\bDelta^b \bD^-\|_\infty
+ \|n^{-1}\bX^{*T}\bE^*\bD^* \bDelta^{d-} \|_\infty,
\end{align*}
where $\bDelta^{d-} = \bD^{-} - \bD^{*-} = \diag\{d_j^{-1}-(d_j^{*})^{-1}\}$. Thus it holds that
\begin{align}
&P\left( \sup_{\bC\in \mathcal{E}_1} \|n^{-1}\bX^{*T}\bE\bB \bD^- \|_\infty \geq \lambda_d \right)
\leq P\left(\|n^{-1}\bX^{*T}\bE^* \|_\infty \geq \lambda_d/3\right) \notag\\
&+ P\left( \sup_{\bC\in \mathcal{E}_1} \|n^{-1}\bX^{*T}\bE\bDelta^b \bD^-\|_\infty \geq \lambda_d/3\right)
+ P\left(  \sup_{\bC\in \mathcal{E}_1}  \|n^{-1}\bX^{*T}\bE^*\bD^* \bDelta^{d-}) \|_\infty \geq \lambda_d/3\right).\label{thm2:csp1:001}
\end{align}

Let us consider the first term on the right hand side of (\ref{thm2:csp1:001}).
Since $\bE \sim N(\bzero,\bI_n \otimes \bSigma)$ by Condition \ref{Aerror},
the $j$th column vector of $\bE^*$, $\be_j^*=\bE\bv_j^*$ with $\bv_j^*$ the $j$th column vector of $\bV^*$,
is distributed as $N\left(0,\bv_j^{*T}\bSigma \bv_j^* I_n\right)$.
Furthermore, note that $\|\bX^{*T}\bE^*\|_\infty=\max_{1\leq i \leq q}\max_{1\leq j \leq q}|\bx_i^{*T}\be_j^*|$ and
\begin{equation}
\mathbb{E} |\bx_i^{*T}\be_j^*|^2 = \bv_j^{*T}\bSigma \bv_j^* \bx_i^{*T}\bx_i^{*}
\leq \alpha_{\max} \bu_i^{*T}\bX^T\bX\bu_i^*
\leq \alpha_{\max}c_3 n\leq cn, \label{thm2:csp1:002}
\end{equation}
where $\alpha_{\max}$ denotes the maximum eigenvalue of $\bSigma$ and
the second inequality follows from Condition \ref{Aeigen} and the fact that $\bu_i^* = \bzero$ for $i = r+1,
\cdots, q$.
Therefore, it follows from Bonferroni's inequality, the Gaussianity of $\be_j^*$, and (\ref{thm2:csp1:002}) that for $\lambda_d^2 = c_1^2 n^{-1}\log (pr)$,
\begin{align}
P\left( n^{-1}\|\bX^{*T}\bE^*\|_\infty \geq \lambda_d/3 \right)
& \leq \sum_{i=1}^r\sum_{j=1}^r P\left( n^{-1}|\bx_i^{*T}\be_j^*| \geq \lambda_d/3 \right) \notag\\
&
\leq 2\sum_{i=1}^r\sum_{j=1}^r \exp \left( -\frac{n^2\lambda_d^2/9}{2\mathbb{E} |\bx_i^{*T}\be_j^*|^2} \right) \notag\\
&\leq 2r^2 \exp \left( -\frac{n^2c_1^2n^{-1}\log (pr)}{18cn} \right) \notag\\
&= 2r^2(pr)^{-c_1^2/c}.\label{thm2:csp1:003}
\end{align}

We now consider the second term on the right hand side of (\ref{thm2:csp1:001}). Some algebra gives
\begin{align}
\|n^{-1}\bX^{*T}\bE\bDelta^b \bD^- \|_\infty
&=\|n^{-1}(\bI_q \otimes \bX^{*T}\bE) \vect(\bDelta^b \bD^-) \|_\infty \notag\\
&\leq \max_{1\leq i\leq r}\sum_{j=1}^q |n^{-1}\bx_i^{*T}\be_j| \|\vect(\bDelta^b \bD^-)\|_\infty \notag\\
&\leq q\max_{1\leq i\leq r}\max_{1\leq j\leq q} |n^{-1}\bx_i^{*T}\be_j| \| (\bD^- \otimes \bI_q)\vect(\bDelta^b )\|_\infty \notag\\
&\leq q\|\bD^{-}\|_\infty\max_{1\leq i\leq r}\max_{1\leq j\leq q} |n^{-1}\bx_i^{*T}\be_j| \|\vect(\bDelta^b)\|_\infty. \notag
\end{align}
Since we solve SOFAR in the local neighborhood $\mathcal{P}_n$ defined in \eqref{def: local-neighbor}, by Condition \ref{Aparsp} we have $\|\bD^{-}\|_\infty \leq \tau^{-1}$ for any $\bC \equiv \bA\bD^{-}\bB \in \mathcal{P}_n$. Thus by \eqref{eq: ABD-bounds}, the second term in the upper bound of (\ref{thm2:csp1:001}) can be bounded as
\begin{align}
\sup_{\bC \in \mathcal{E}_1}\|n^{-1}\bX^{*T}\bE\bDelta^b \bD^- \|_\infty
&\leq (q/\tau)\max_{1\leq i\leq r}\max_{1\leq j\leq q} |n^{-1}\bx_i^{*T}\be_j| \sup_{\mathcal{E}_1}\|\vect(\bDelta^b)\|_\infty \notag\\
&\leq (q/\tau)\max_{1\leq i\leq r}\max_{1\leq j\leq q} |n^{-1}\bx_i^{*T}\be_j| \sup_{\mathcal{E}_1}\|\bDelta^b\|_F \notag\\
&\leq 3c (q/\tau)\eta_n R_n \max_{1\leq i\leq r}\max_{1\leq j\leq q} |n^{-1}\bx_i^{*T}\be_j|\label{thm2:csp1:004}.
\end{align}
Similarly to (\ref{thm2:csp1:002}), we can show that
\begin{align}
\mathbb{E} |\bx_i^{*T}\be_j|^2
\leq \sigma_j^2 c_3 n \leq \sigma_{\max}^2 c_3n\leq cn. \label{thm2:csp1:005}
\end{align}
Therefore, in view of (\ref{thm2:csp1:004}), (\ref{thm2:csp1:005}), $R_n^2= O(sn^{-1}\log (pq))$, and 
$p \geq q$,
the same inequality as (\ref{thm2:csp1:003}) results in
\begin{align}
&P\left( \sup_{\mathcal{E}_1}\|n^{-1}\bX^{*T}\bE\bDelta^b \bD^- \|_\infty \geq \lambda_d/3\right) \notag\\
&=P\left( 3c (q/\tau) \eta_nR_n \max_{1\leq i\leq r}\max_{1\leq j\leq q} |n^{-1}\bx_i^{*T}\be_j| \geq \lambda_d/3\right) \notag\\
&\leq 2\sum_{i=1}^r\sum_{j=1}^q\exp\left( -\frac{n^2\lambda_d^2}{81c(q/\tau)^2\eta_n^2R_n^2\mathbb{E} |\bx_i^{*T}\be_j|^2} \right) \notag\\
&= 2qr \exp\left( -\frac{c_1^2n}{c(q/\tau)^2\eta_n^2s} \right),\label{thm2:csp1:006}
\end{align}
where $c$ is some 
positive constant.

It remains to investigate the third term on the right hand side of (\ref{thm2:csp1:001}).
Since $\bD^*\bDelta^{d-}$ is a diagonal matrix whose $(k,k)$th entry is given by
$(d_k^*-d_k)/d_k$
with $\rank(\bD^*\bDelta^{d-})\leq r$, the last $q-r$ columns of both $\bX^*$ and $\bE^*$ are zero, and 
$\bD \in \mathcal{D}$, we have
\begin{align}
\sup_{\bC \in \mathcal{E}_1} \|n^{-1}\bX^{*T}\bE^*\bD^* \bDelta^{d-} \|_\infty
&\leq \max_{1\leq i \leq r} \max_{1\leq j \leq r} |n^{-1}\bx_i^{*T}\be_j^*| \sup_{\mathcal{E}} \| \bD^* \bDelta^{d-}\|_\infty \notag\\
&\leq \tau^{-1}\max_{1\leq i \leq r} \max_{1\leq j \leq r} |n^{-1}\bx_i^{*T}\be_j^*| \max_{1\leq k\leq r} |d_k^*-d_k| \notag\\
&\leq \tau^{-1}\max_{1\leq i \leq r} \max_{1\leq j \leq r} |n^{-1}\bx_i^{*T}\be_j^*| \|\bDelta^d\|_F \notag\\
&\leq 3(R_n/\tau) \max_{1\leq i \leq r} \max_{1\leq j \leq r} |n^{-1}\bx_i^{*T}\be_j^*|. \label{thm2:csp1:007}
\end{align}
Then by (\ref{thm2:csp1:002}) and (\ref{thm2:csp1:007}), the same inequality yields
\begin{align}
P\left( \sup \|n^{-1}\bX^{*T}\bE^*\bD^* \bDelta^{d-} \|_\infty \geq \lambda_d/3\right)
&\leq P\left( 3(R_n/\tau) \max_{1\leq i\leq r}\max_{1\leq j\leq r} |n^{-1}\bx_i^{*T}\be_j^*| \geq \lambda_d/3\right) \notag\\
&\leq 2\sum_{i=1}^r\sum_{j=1}^r\exp\left( -\frac{n^2\lambda_d^2}{81c(R_n/\tau)^2\mathbb{E} |\bx_i^{*T}\be_j^*|^2} \right) \notag\\
&\leq 2r^2\exp\left( -\frac{c_1^2n^2n^{-1}\log(pr)}{csn^{-1}\log(pq)\tau^{-2}n} \right) \notag\\
&\leq 2r^2\exp\left( -\frac{c_1^2\tau^2n}{cs} \right). \label{thm2:csp1:008}
\end{align}
Therefore, combining (\ref{thm2:csp1:003}), (\ref{thm2:csp1:006}), and (\ref{thm2:csp1:008}) with (\ref{thm2:csp1:001}) gives the probability
bound
\begin{align}
&P\left( \sup_{\bC\in\mathcal{E}_1} \|n^{-1}\bX^{*T}\bE\bB \bD^- \|_\infty \geq \lambda_d \right) \notag \\
&\leq 2r^2(pr)^{-c_1^2/c} + 2rq \exp\left( -\frac{c_1^2n}{c(q/\tau)^2\eta_n^2s} \right) + 2r^2\exp\left( -\frac{c_1^2\tau^2n}{cs} \right).
\label{thm2:csp1:009}
\end{align}

We next prove that (\ref{A:csp2-new}) holds with high probability. The arguments are similar to those for proving (\ref{A:csp1-new}) except that $\bX^*$ is replaced with $\bX$ in the proof of (\ref{A:csp1}).
More specifically, note that we have the following decomposition of probability bound
\begin{align}
&P\left( \sup_{\mathcal{E}_1} \|n^{-1}\bX^{T}\bE\bB \bD^- \|_\infty \geq \lambda_a \right)
\leq P\left(\|n^{-1}\bX^{T}\bE^* \|_\infty \geq \lambda_a/3\right) \label{thm2:csp2:001}\\
&+ P\left( \sup_{\mathcal{E}_1} \|n^{-1}\bX^{T}\bE\bDelta^b \bD^-\|_\infty \geq \lambda_a/3\right)
+ P\left( \sup_{\mathcal{E}_1} \|n^{-1}\bX^{T}\bE^*\bD^* \bDelta^{d-}) \|_\infty \geq \lambda_a/3\right). \notag
\end{align}
Thus, it suffices to bound the probabilities on the right hand side of (\ref{thm2:csp2:001}).
Let us consider the first term. Observe that
\begin{align*}
\mathbb{E} |\bx_i^{T}\be_j^*|^2
\leq \alpha_{\max} \bx_i^T\bx_i = \alpha_{\max}n \leq cn,
\end{align*}
where $c$ is some 
positive constant.
Thus, setting
$\lambda_a^2 = c_1^2 n^{-1}\log (pr)$ and noting that $\bE^*$ has only $r$ nonzero columns
lead to the bound
\begin{align}
P\left( n^{-1}\|\bX^{T}\bE^*\|_\infty \geq \lambda_a/3 \right)
&\leq 2\sum_{i=1}^p\sum_{j=1}^r \exp \left( -\frac{n^2\lambda_a^2}{8\mathbb{E} |\bx_i^{T}\be_j^*|^2} \right) \notag\\
&\leq 2pr \exp \left( -\frac{c_1^2 n^2n^{-1}\log (pr)}{cn} \right) \notag\\
&\leq 2(pr)^{1-c_1^2/c}. \label{thm2:csp2:002}
\end{align}
We next consider the second probability bound on the right hand side of (\ref{thm2:csp2:001}). Since
\begin{align*}
\mathbb{E} |\bx_i^{T}\be_j|^2
\leq \sigma_{\max}^2 \bx_i^T\bx_i = \sigma_{\max}^2 n \leq cn,
\end{align*}
by replacing $\max_{1\leq i \leq r}$ in (\ref{thm2:csp1:004}) and (\ref{thm2:csp1:006})
with $\max_{1\leq i \leq p}$ we deduce
\begin{align}
P\left( \sup_{\bC \in \mathcal{E}_1} \|n^{-1}\bX^{T}\bE\bDelta^b \bD^-\|_\infty \geq \lambda_a/3\right)
&\leq 2pr \exp\left( -\frac{c_1^2n}{c(q/\tau)^2\eta_n^2s} \right). \label{thm2:csp2:003}
\end{align}
It remains to study the third probability bound on the right hand side of (\ref{thm2:csp2:001}).
Similarly, replacing $\max_{1\leq i \leq r}$ in (\ref{thm2:csp1:007}) and (\ref{thm2:csp1:008})
with $\max_{1\leq i \leq p}$ yields
\begin{align}
P\left( \sup_{\bC\in \mathcal{E}_1} \|n^{-1}\bX^{T}\bE^*\bD^* \bDelta^{d-} \|_\infty \geq \lambda_a/3\right)
&\leq  2pr\exp\left( -\frac{c_1^2\tau^2n}{cs} \right). \label{thm2:csp2:004}
\end{align}
Thus combining (\ref{thm2:csp2:002})--(\ref{thm2:csp2:004}), we can bound (\ref{thm2:csp2:001}) as
\begin{align}
&P\left( \sup_{\bC\in \mathcal{E}_1}  \|n^{-1}\bX^{T}\bE\bB \bD^- \|_\infty \geq \lambda_a \right)  \notag \\
&\leq
2(pr)^{1-c_1^2/c}
+ 2pr \exp\left( -\frac{c_1^2n}{c(q/\tau)^2\eta_n^2s} \right)
+ 2pr\exp\left( -\frac{c_1^2\tau^2n}{cs} \right). \label{thm2:csp2:005}
\end{align}

Finally, we show that condition (\ref{A:csp3-new}) holds with large probability.
Choosing $\lambda_b^2 = c_1^2 n^{-1}\log (pr)$ results in
\begin{align}
P\left( n^{-1}\|\bX^{*T}\bE\|_\infty \geq \lambda_b \right)
&\leq 2\sum_{i=1}^r\sum_{j=1}^q \exp \left( -\frac{n^2\lambda_b^2}{2\mathbb{E} |\bx_i^{*T}\be_j|^2} \right) \notag\\
&\leq 2qr \exp \left( -\frac{c_1^2n^2n^{-1}\log(pr)}{cn} \right) \notag\\
&\leq 2qr(pr)^{-c_1^2/c}. \label{thm2:csp3:001}
\end{align}
Consequently, for the given set of regularization parameters $(\lambda_d, \lambda_a, \lambda_b)$ it follows from (\ref{thm2:csp1:009}), (\ref{thm2:csp2:005}), and (\ref{thm2:csp3:001})  
that conditions (\ref{A:csp1-new})--(\ref{A:csp3-new}) hold simultaneously with probability at least
\begin{align*}
& 1 -
\left\{
2(pr)^{1-c_1^2/c}
+ 2pr \exp\left( -\frac{c_1^2n}{c(q/\tau)^2\eta_n^2s} \right)
\right\},
\end{align*}
where we have used the facts of $c_1^2>c$ and $p\geq q \geq 1$. 
Moreover, to check that the probability bound converges to one,
since $c_1^2>c$ it is sufficient to show that
\begin{align*}
2pr \exp\left( -\frac{c_1^2n}{c(q/\tau)^2\eta_n^2s} \right)
\end{align*}
converges to zero. This follows immediately from the assumptions of $\log p = O(n^{\alpha})$, $q = O(n^{\beta/2})$, $s = O(n^{\gamma})$, and $\eta_n/\tau = o(n^{(1-\alpha-\beta-\gamma)/2})$, which concludes the proof of Lemma \ref{lem:inequalities}.

\subsection{Proof of Theorem \ref{thm:conv}} \label{secA.3}
Recall that the theoretical results for the SOFAR estimator established in the paper hold simultaneously over the set of all local minimizers in a neighborhood of the initial Lasso estimator. Thus we aim to establish the convergence of the SOFAR algorithm when supplied the initial Lasso estimator. Note that the equivalent form of the SOFAR problem (\ref{eq:sofar1}) with the slack variables $\bA$ and $\bB$ can be solved using the augmented Lagrangian form with sufficiently large penalty parameter $\mu > 0$. From now on, we fix parameter $\mu$ and the set of Lagrangian multipliers $\bGamma$, and thus work with the objective function $L_\mu(\bTheta, \bOmega; \bGamma)$.

By the nature of the block coordinate descent algorithm applied to $(\bU, \bV, \bD, \bA, \bB)$, the sequence $(L_\mu(\cdot))$ of values of the objective function $L_\mu(\bTheta, \bOmega; \bGamma)$ is decreasing. Clearly the function $L_\mu(\bTheta, \bOmega; \bGamma)$ is bounded from below. Thus the sequence $(L_\mu(\cdot))$ converges. Since the rank parameter $m$ is fixed in the SOFAR algorithm, we assume for simplicity that the diagonal matrix $\bD^k$ of singular values has all the diagonal entries bounded away from zero, since otherwise we can solve the SOFAR problem with a smaller rank $m$.

By assumption, we have
\[ \sum_{k = 1}^\infty [\Delta L_\mu(\bU^k)]^{1/2} < \infty, \ \sum_{k = 1}^\infty [\Delta L_\mu(\bV^k)]^{1/2} < \infty, \ \text{ and } \ \sum_{k = 1}^\infty [\Delta L_\mu(\bD^k)]^{1/2} < \infty, \] where $\Delta L_\mu(\cdot)$ stands for the decrease in $L_\mu(\cdot)$ by a block update. Note that the $\bU$-space with constraint $\bU^T\bU=\bI_m$ is a Stiefel manifold which is compact and smooth; see, e.g., \cite{Lv2013} for a brief review of the geometry of Stiefel manifold. Since the $\bD$-sequence is always positive definite by assumption, the objective function along the $\bU$-block with all the other four blocks fixed is convex and has positive curvature bounded away from zero along any direction in the $\bU$-space. By definition, $\bU^{k}$ is the minimizer of such a restricted objective function, which entails that the gradient of this function at $\bU^{k}$ on the Stiefel manifold vanishes. Thus it follows easily from the mean value theorem and the fact of positive curvature that $\Delta L_\mu(\bU^k)$ is bounded from below by some positive constant $\delta$ times $d_g^2(\bU^{k}, \bU^{k-1})$, where $d_g(
\cdot, \cdot)$ denotes the distance function on the Stiefel manifold. Then it holds that
\[ \sum_{k = 1}^\infty d_g(\bU^{k}, \bU^{k-1}) \leq
\delta^{-1/2} \sum_{k = 1}^\infty [\Delta L_\mu(\bU^k)]^{1/2} < \infty, \]
which along with the triangle inequality entails that $(\bU^k)$ is a Cauchy sequence on the Stiefel manifold. Therefore, the sequence $(\bU^k)$ converges to a limit point $\bU_*$ on the Stiefel manifold which is a local solution along the $\bU$-block. Similarly, we can show that the sequence $(\bV^k)$ also converges to a limit point $\bV_*$ on the Stiefel manifold that is a local solution along the $\bV$-block.

Recall that the diagonal matrix $\bD^k$ of singular values is assumed to have all the diagonal entries bounded away from zero. Since we have shown that the sequences $(\bU^k)$ and $(\bV^k)$ converge to limit points $\bU_*$ and $\bV_*$ on the Stiefel manifolds, respectively, it follows from the fact that both $\bU_*$ and $\bV_*$ have full column rank $m$ that as $k$ becomes large, the objective function along the $\bD$-block with all the other four blocks fixed is convex and has positive curvature bounded away from zero. Thus an application of similar arguments as above yields that the sequence $(\bD^k)$ also converges to a limit point $\bD_*$.

With the established convergence results of the sequences $(\bU^k)$, $(\bV^k)$, and $(\bD^k)$, the convergence of the sequences $(\bA^k)$ and $(\bB^k)$ follows easily from the convergence property of the block coordinate descent algorithm applied to separable convex problems \citep{Tsen:conv:2001}, by noting that the objective function with $\bU$, $\bV$, and $\bD$ replaced by their limit points is jointly convex in $\bA$ and $\bB$ since the penalty functions $\rho_a(\cdot)$ and $\rho_b(\cdot)$ are assumed to be convex. This completes the proof of Theorem \ref{thm:conv}.

\section{Additional technical details} \label{secB}

\subsection{Lemma \ref{lem:MPT} and its proof} \label{secB.3}

\begin{lemma}\label{lem:MPT}
Under Condition \ref{Asgap}, we have for any matrix $\bC=\bU\bD\bV^T$ and $\bC^*=\bU^*\bD^*\bV^{*T}$ with $\|\bC-\bC^*\|_2\leq d_1^*$ that
\begin{align*}
\|\bD-\bD^*\|_F &\leq \|\bC-\bC^*\|_F, \\
\|\bA-\bA^*\|_F + \|\bB-\bB^*\|_F &\leq c \eta_n \|\bC-\bC^*\|_F,
\end{align*}
where $\eta_n=1+\delta^{-1/2} \big(\sum_{j=1}^r(d_1^*/d_j^*)^2\big)^{1/2}$ and $c>0$ is some constant.
\end{lemma}

\textit{Proof of Lemma \ref{lem:MPT}}. It is well known that the inequality \[ \|\bD-\bD^*\|_F \leq \|\bC-\bC^*\|_F \]
holds; see, for example, \cite{Mirsky1960}. It remains to show the second desired inequality. Recall that $\bA^*=\bU^*\bD^*$.
By the decomposition
\begin{align*}
\bC-\bC^* = (\bA-\bA^*)\bV^T + \bA^*(\bV-\bV^*)^T
\end{align*}
and the unitary property of the Frobenius norm, we have
\begin{align}
\|\bA-\bA^*\|_F \leq \|\bC-\bC^*\|_F + \|\bD^*(\bV-\bV^*)^T\|_F. \label{lem:perturb1}
\end{align}
Let us examine the second term on the right hand side of (\ref{lem:perturb1}). To do so, we apply Theorem 3 of \cite{Yu:Wang:Samworth:2015} to $\bV-\bV^*$ columnwise to avoid the identifiability issue.  When $r=1$ or 2, it holds that
\begin{align}
\|\bv_1-\bv_1^*\|_2
\leq \frac{cd_1^*\|\bC-\bC^*\|_F}
{\delta^{1/2}(d_{1}^{*})^2}, ~~~~
\|\bv_r-\bv_r^*\|_2
\leq \frac{cd_1^*\|\bC-\bC^*\|_F}
{\delta^{1/2}(d_{r}^{*})^2}. \label{eq: vineq-1}
\end{align}
When $r\geq 3$, in addition to \eqref{eq: vineq-1} we have for $j=2,\dots,r-1$,
\begin{align}
\|\bv_j-\bv_j^*\|_2 &\leq \frac{c(2d_1^*+\|\bC-\bC^*\|_2)\|\bC-\bC^*\|_F}
{\min(d_{j-1}^{*2}-d_j^{*2},d_j^{*2}-d_{j+1}^{*2})},  \notag
\end{align}
where $c>0$ is some 
constant.
Since Condition \ref{Asgap} gives $d_{j-1}^{*2}-d_j^{*2}\geq \delta^{1/2}(d_{j-1}^*)^2 \geq \delta^{1/2}(d_{j}^*)^2$, it follows from the assumption $\|\bC-\bC^*\|_2\leq d_1^*$ that the above inequality can be further bounded as
\begin{align*}
\|\bv_j-\bv_j^*\|_2 &\leq \frac{c(2d_1^*+\|\bC-\bC^*\|_2)\|\bC-\bC^*\|_F}
{\min(d_{j-1}^{*2}-d_j^{*2},d_j^{*2}-d_{j+1}^{*2})} \leq \frac{cd_1^*\|\bC-\bC^*\|_F}
{\delta^{1/2}(d_j^*)^2}.
\end{align*}
Thus these inequalities entail that
\begin{align}
\|\bD^* (\bV-\bV^*)^T\|_F^2 = \sum_{j=1}^r d_j^{*2} \|\bv_j-\bv_j^*\|_2^2 \leq  (c/\delta)\|\bC-\bC^*\|_F^2\sum_{j=1}^r(d_1^{*}/d_j^*)^2.\label{lem:perturb5}
\end{align}

Consequently, combining (\ref{lem:perturb1}) and (\ref{lem:perturb5}) leads to the bound
\begin{align*}
\|\bA-\bA^*\|_F \leq  \|\bC-\bC^*\|_F + (c/\delta^{1/2})\|\bC-\bC^*\|_F \left\{\sum_{j=1}^r(d_1^{*}/d_j^*)^2\right\}^{1/2}. 
\end{align*}
On the other hand, the bound for $\|\bB-\bB^*\|_F$ can be obtained by the decomposition
$\bC-\bC^* = \bU(\bB-\bB^*)^T + (\bU-\bU^*)\bB^{*T}$ and similar arguments. Therefore, adding both bounds together and enlarging the positive constant $c$ conclude the proof of Lemma \ref{lem:MPT}.

\subsection{Lemma \ref{lem:rsc} and its proof} \label{secB.1}


\begin{lemma} \label{lem:rsc}
Under Conditions \ref{Aparsp} and \ref{Aeigen}, it holds for any $\bC \in \mathcal{C}$ that
\begin{align*}
n^{-1} \|\bX(\bC-\bC^*)\|_F^2 \geq c_2 \|\bC-\bC^* \|_F^2.
\end{align*}
\end{lemma}

\textit{Proof of Lemma \ref{lem:rsc}}. Denote by $\bDelta = \bC - \bC^*$, $\bW=\bI_q \otimes \bX$, and $\bdelta=\vvec(\bDelta)$, where $\bI_q$ is the $q \times q$
identity matrix. It follows from the triangle inequality and Condition \ref{Aparsp} that
\begin{align*}
\|\bdelta\|_0 & = \|\vvec(\bC) - \vvec(\bC^*)\|_0 \leq \|\vvec(\bC)\|_0 + \|\vvec(\bC^*)\|_0 \\
& < \kappa_{c_2}/2 + \kappa_{c_2}/2 = \kappa_{c_2}.
\end{align*}
Note that the singular values of $\bW$ are the same as those of the original design matrix $\bX$ with the multiplicity of
each singular value multiplied by $q$. This entails that the robust spark of $\bW$ is equal to that of $\bX$, which is $\kappa_{c_2}$
for a given positive constant $c_2$. Thus by the definition of the robust spark, we obtain
\begin{align*}
n^{-1} \|\bX\bDelta\|_F^2
= n^{-1} \|\bW\bdelta\|_2^2
= n^{-1} \|\bW_{\supp(\bdelta)} \bdelta_{\supp(\bdelta)} \|_2^2
\geq c_2 \|\bdelta \|_2^2
= c_2 \|\bDelta \|_F^2,
\end{align*}
where the subscript $\supp(\bdelta)$ denotes the restriction of the matrix to the corresponding columns or that of the vector to the corresponding components. This completes the proof of Lemma \ref{lem:rsc}.

\end{document}